\documentclass{pasj00}                
\draft

\author{Hiroyuki \textsc{Yoshiguchi}$^1$, Shigehiro \textsc{Nagataki}$^1$, Katsuhiko \textsc{Sato}$^{1,3}$,\\ Nobuyuki \textsc{Ohama}$^2$ and Sadanori \textsc{Okamura}$^{2,3}$}
\affil{$^1$ Department of Physics, School of Science, the University
of Tokyo, \\ 7-3-1 Hongo, Bunkyoku, Tokyo 113-0033 }
\affil{$^2$ Department of Astronomy, School of Science, the University
of Tokyo, \\ 7-3-1 Hongo, Bunkyoku, Tokyo 113-0033 }
\affil{$^3$ Research Center for the Early Universe, School of
Science, the University of Tokyo, \\ 7-3-1 Hongo, Bunkyoku, Tokyo 113-0033,
Japan\\}

\email{hiroyuki@utap.phys.s.u-tokyo.ac.jp}
\title{Revision of the Selection Function \\
of the Optical Redshift Survey\\
using the Sloan Digital Sky Survey Early Data Release:\\
Toward an Accurate Estimate of\\ Source Number Density of\\
Ultra-High Energy Cosmic Rays}
\Received{}
\Accepted{}

\begin{document}

\maketitle
\begin{abstract}
If Ultra-High Energy Cosmic Rays (UHECRs) are originated from nearby
galaxies, modeling of the distribution of nearby galaxies is important
to an accurate estimate the source number density of UHECRs.
We investigate uncertainty of the selection function of the Optical
Redshift Survey (ORS), which we used to construct a source model of
UHECRs.
The investigation is based on a comparison of numbe counts of
ORS galaxies with those of the spectroscopic sample of the Sloan Digital Sky
Survey (SDSS) Early Data Release (EDR).
We carefully count galaxies in the same absolute magnitude bin from the
two samples.
We find a slight systematic overestimate of the ORS 
counts outside 5000 km s$^{-1}$ by about a factor of 2.
We revise the selection function of the ORS assuming that the SDSS counts
are correct.
Our revision is based on the absorption given in the ORS catalog
as well as that computed from Schlegel et al. (1998), which is systematically 
larger than the former by $A_B \sim 0.1$ mag in the region of low absorption.
It is found that introduction of Schlegel et al.'s absorption changes
one of the parameters of the ORS selection function by more than $10 \%$.
The revision should be taken into account in the future analysis
of the source number density of UHECRs based on the ORS.
Using the revised selection function, we determine the global structure of 
the Local Supercluster (LSC) with a source model of UHECRs, that is,
a number-density model consisting of a uniform spherical halo and
an exponential disk.
We find that the revision is insignificant in terms of the structure 
of the LSC.
However, the revised selection function will be useful to
other studies such as peculiar velocity and correlation function.

\end{abstract}

\section {INTRODUCTION} \label{intro}
\indent

Since galaxies are tracers of the underlying matter distribution
in the universe, they are often used as sources of
Ultra High Energy Cosmic Rays (UHECRs), i.e., cosmic rays with energy
above 10$^{19.5}$ eV,
in conventional acceleration scenarios (Ide et al. 2001;
Blanton et al. 2001; Smialkowski et al. 2002; Yoshiguchi et al. 2002).
In such scenarios UHECRs are considered to be charged particles.
Sources of UHECRs must be located in the limited volume
within a typical radius $\sim$ 50 Mpc
because of interaction of such particles with CMB photons 
(GZK effect) (Greisen 1966; Zatsepin and Kuz`min 1966).
On the other hand, cosmic rays below 10$^{19.5}$ eV
are able to come from much larger volume
of radius $\sim$ 100 Mpc, which is determined
by deflection in the intergalactic magnetic field.
The cosmic ray spectrum should therefore show a sharp cut off at
10$^{19.5}$ eV (GZK cutoff).
However, the observed spectrum is featureless and
the GZK cutoff can not be seen (Takeda et al. 1999).
There is another problem concerning arrival directions of UHECRs.
No significant large scale anisotropy is found in arrival directions
of 47 UHECRs observed so far.
However, one triplet and three doublets with a separation angle of
$2.5^{\circ}$ are observed among the 47 events.
Chance probability to observe such clustered events under an isotropic
distribution is smaller than 1 $\%$ (Takeda et al. 1999).

In order to resolve these problems, there are several groups
which perform numerical simulation for the energy spectrum and
arrival directions of UHECRs.
Previous studies can be divided roughly into two types.
The first uses actual galaxy samples as realistic source models
of UHECRs as mentioned above (Ide et al. 2001; Blanton et al. 2001;
Smialkowski et al. 2002; Yoshiguchi et al. 2002).
The second adopts simple source models which represent the Local
Supercluster (LSC), such as a two dimensional
sheet (Sigl et al. 1999) or
a highly flattened oblate spheroid (Lemoine et al. 1999).

A major difficulty in calculating the UHECRs flux numerically
with an actual galaxy sample is that galaxies in a given magnitude-limited
sample are biased tracers of matter distribution because of the flux-limit.
In our previous study, we examined whether current observations of UHECRs
could be explained by conventional acceleration scenarios (Ide et al. 2001;
Yoshiguchi et al. 2002)
using a whole sky survey, Optical Redshift Survey (ORS; Santiago et al. 1995).
However, the correction for the selection effect is often very large.
For example, the selection function of the ORS given
in Santiago et al.(1996) is roughly equal to 0.01 at 8000 km s$^{-1}$.
Blanton et al.(2001) showed that the small local overdensity which
they derived from the Center for Astrophysics Redshift Survey and
the IRAS PSCz Survey
gave the predicted GZK cutoff which is discrepant with the current
observation at only 2 $\sigma$ level.
As is evident from this result,
it is important to know uncertainty of the large correction.

A major concern with studies of the second type is that their source
models are too primitive.
It should be emphasized that UHECRs above 10$^{19.5}$ eV must
originate within about 50 Mpc from us.
Accordingly, the spatial distribution of nearby galaxies plays an
important role for the distribution of arrival directions of UHECRs.
However, in addition to the primitive form of their models,
parameters such as radial and vertical
length of the models are not well constrained.

In this paper, we investigate the uncertainty of the ORS selection function
by comparing the galaxy number counts of galaxies
as a function of recession velocity
between Sloan Digital Sky Survey: Early Data Release 
(SDSS EDR) (Stoughton et al. 2002)
and the ORS.
The SDSS is the imaging and spectroscopic
survey in five optical frequency bands
with a limiting magnitude which is much deeper than that of the ORS,
but covering a very limited area of the ORS.

The plan of this paper is as follows:
We introduce the SDSS and ORS data in Section \ref{data}.
Effect of Galactic absorption is described in Section \ref{obsc}.
Comparison of the galaxy number counts and
revised parameter of the ORS selection function
are given in Section \ref{calibration}.
In Section \ref{model}, we describe a source model of UHECRs, that is,
a model of the galaxy number density, 
the method of determining its parameters, and resulting parameters.
In Section \ref{summary}, we summarize the results and discuss
their implications.

\indent
\section{THE GALAXY SAMPLES} \label{data}
\indent

\subsection{THE ORS SAMPLE} \label{ors}
\indent

The ORS is drawn from three catalogs,
the European Southern Observatory Galaxy Catalogue (ESO)
covering the southern sky ($\delta \leq -17.5^{\circ}$),
the Uppsala Galaxy Catalogue (UGC)
in the northern sky ($\delta  \geq -2.5^{\circ}$),
and the Extension to the Southern Galaxy Catalogue (ESGC)
in the strip between the ESO and UGC region
($-17.5^{\circ} \leq \delta  \leq -2.5^{\circ}$).
The last catalogue contains no magnitudes
other than very poor ones from eye inspection.
The sample selection and galaxy distribution
are described in detail by Santiago et al. (1995).
This survey contains two subcatalogs, one complete to a B magnitude
of 14.5 (ORS-m) and the other complete to a B major axis diameter of $1.'9$
(ORS-d).
In this study we calibrate the ORS selection function using the SDSS
EDR which is a magnitude-limited survey. Accordingly, we use
the magnitude-limited subcatalog, i.e., ORS-m, which
contain 5716 galaxies.
The subcatalog ORS-m itself is also composed 
of two subsamples, ESO-m (2437 galaxies) and UGC-m (3279 galaxies). 
We further restrict ourselves to galaxies within 8000 km s$^{-1}$
and with $M_B - 5$log$_{10}h < -0.5 - 5$log$_{10}v$, which corresponds to
the apparent magnitude limit of $m_B=14.5$,
where $h$ is the dimensionless Hubble constant
$h=H_0/(100$ km s$^{-1}$ Mpc$^{-1})$ and $v$ is the recession velocity.
Very faint nearby galaxies with $M_B>-14.0$ are excluded.
Galaxies in the zone of avoidance ($|b|<20^{\circ}$) are also excluded. 
Our final ORS sample contains 5178 galaxies with 2346 galaxies
in the ESO-m and 2832 galaxies in the UGC-m.

Figure~\ref{fig1} shows the sky distribution of our ORS galaxies
in the equatorial coordinate. Figure~\ref{fig2} shows the absolute magnitude
plotted as a function of the recession velocity.
Throughout the paper, we assume that there are no departures from
uniform Hubble expansion for simplicity.
We use $h$ = 1 when we do not specify $h$ explicitly.
An alternative value for $h$ changes all absolute magnitudes by 5log$_{10} h$.


In order to compare the galaxy number counts of the ORS
with those of the SDSS EDR,
a selection function has to be introduced.
This selection function of the ORS is derived
for several subsamples in Santiago et al.(1996)
using galaxies to 8000 km s$^{-1}$.
They assumed that the selection function is expressed as
\begin{equation}
\phi_{\rm ORS}(r) = \left \{
\begin{array}{ll}
\left( \frac{r}{r_{s}}\right) ^{-2a}
\left( \frac{r_{*}^2 + r^2}{r_{*}^2 + r_{s}^2} \right)
^{-b} &\quad r > r_{s} \\
1 &\quad r \leq r_{s}
\end{array}
\right.,
\end{equation}
where $r$ represents distance from our Galaxy,
and they set $r_{s}$ to be 500 km s$^{-1}$.
Values of the parameters determined by Santiago et al.(1996)
are $a$ = 0.40, $b$ = 6.25, $r_{*}$ = 11100 (km s$^{-1}$) for ESO-m
and $a$ = 0.36, $b$ = 9.38, $r_{*}$ = 14060 (km s$^{-1}$) for UGC-m.
The flux limit $m_B = 14.5$ corresponds to
$M_B = -14$ at $v = 500$ km s$^{-1}$,
where the selection function is unity.
In Figure~\ref{fig3} we show the selection function of the ORS
given by Santiago et al.(1996) as a function of recession velocity
together with that of the SDSS (explained in the next subsection).


\indent
\subsection{THE SDSS SAMPLE} \label{sdss}
\indent

\subsubsection{Completeness} \label{complete}
\indent

The SDSS is the imaging survey in five bands ($u', g', r', i', z'$)
followed by spectroscopy. The spectroscopic sample of the SDSS EDR
contains 1504 galaxies with measured recession velocity $v<8000$ km s$^{-1}$.
The region covered by the spectroscopy of the SDSS EDR is 
shown in Figure~\ref{fig4}.
In the following, we refer to the region at 
$\delta>50^{\circ}$ as region 1, two regions in the equatorial
stripe as regions 2 and 3.
All the data in these three regions are collectively
referred to as the ALL region.


At the brightest end, it is suspected that problems in the
deblender and/or possible saturation limit the completeness of the SDSS.
There are 130 ORS galaxies which fall in the ALL region.
We tried to match these 130 galaxies with the SDSS catalogs.
We found that 72 ORS galaxies out of the 130 could be matched
with the SDSS spectroscopic objects and that
43 could be matched with the photometric objects
(but not with the spectroscopic objects).
Remaining 15 galaxies are not included either in the spectroscopic catalog
nor in the photometric catalog.
Among them, 10 suffer from problems in the deblender, 1 is too bright
($\sim$ 9 mag), and remaining 4 galaxies are missed by some unknown reason.
Thus the completeness of the SDSS spectroscopic catalog is quite low,
$72/130=55\%$, at the ORS limiting magnitude.
However, if supplemented by the photometric catalog, it increases to
$115/130=88\%$.
The SDSS spectroscopic catalog have a very high completeness for
galaxies including very faint ones (above $99\%$ for $r'<18$)
(Strauss et al. 2002).
We add the 43 SDSS photometric objects in the above
1504 galaxies assigning them the recession velocities given in the ORS.
Thus, our SDSS sample now includes 1547 galaxies.
We extract our final sample from these 1547 galaxies to be compared
with the ORS sample including the 130 galaxies.

\indent
\subsubsection{Conversion from $m_B$ to $g'$ magnitude} \label{conversion}
\indent

The ORS B magnitude is based on the photographic plates and its
bandpass is similar to that of Couch-Newell B$_J$ (Couch and Newell 1980).
We use only SDSS $g'$-band magnitude in this study because
it is close to the B$_J$ magnitude.
In order to carefully count galaxies in the same magnitude range,
we have to know magnitude difference between $m_B$ and $g'$.
For this purpose, we made a direct comparison of the magnitudes
of the matched 72 plus 43 galaxies between the two surveys.
However, variance of the magnitude offset was very large
($g'-m_B \sim 1.5$ at the maximum).
At such bright end (the ORS-m is complete to $m_B=14.5$ mag),
it is suspected that problems in the deblender limit
the accuracy of the SDSS magnitude. 
Accordingly, we derive the magnitude offset in the method explained below.

There are three differences of the magnitude system
between the SDSS and the ORS.
First, response functions of these two photometric band systems differ
from each other.
Magnitude offset due to this difference can be computed from galaxy colors in
various photometric band systems given in Fukugita et al.(1995) for typical
morphology types grouped into E, S0, Sa-Sb, Sb-Sc, Sc-Sd, and Sdm-Im.
The ORS gives the morphological type index to individual galaxy,
whereas the SDSS does not.
However, Strateva et al.(2001) showed that galaxies have a bimodal $u^*-r^*$
color distribution corresponding to early (E, S0, Sa) and late (Sb, Sc, Irr)
morphological types and that the two types can be clearly separated
by a $u^*-r^*$ color cut of 2.22 independent of magnitude.
Accordingly, we compute the magnitude offset for early and late
morphological types and use them throughout the paper.
Averaging over morphology types of galaxies in our ORS sample,
we adopt $g'-B_J=-0.35$ and $-0.24$ for early and late type galaxies,
respectively.

Second, there is a difference about the method for measuring the flux.
In order to measure a constant fraction of the total light,
independent of Galactic absorption and cosmological dimming,
the SDSS has adopted a modified form of the Petrosian (1976) magnitude.
Magnitudes are measured within the aperture of twice the Petrosian
radius, which is insensitive to absorption or cosmological dimming.
The detailed explanations are presented in Blanton et al.(2001),
Yasuda et al.(2001), and Stoughton et al.(2002).
On the other hand, the ORS is based on an isophotal magnitude.
The relationship between the SDSS Petrosian magnitudes and isophotal
magnitudes is examined in Blanton et al.(2001).
Difference between the two magnitudes 
$\Delta m = m_{petro}-m_{iso}$ ranges from $-0.2$ to $0.1$
according to several choices of the isophotal limit in case of $r'$ band.
It is, however, not known exactly what value we should take for $\Delta m$
in the $g'$ band.
Here we tentatively take $\Delta m = -0.1$ with an estimated uncertainty
of $\pm 0.2$.

Finally, SDSS magnitudes are not the traditional logarithmic
magnitude but inverse hyperbolic sine magnitude,
which is described in detail by Lupton et al.(1999).
However, these magnitudes differ by less than 1 \% in flux above $g'=22.60$.
We neglect this difference throughout the paper.

Taking account of these three differences
between the SDSS and the ORS magnitudes, we take the total amount of the
magnitude offset as $g'-B_J+\Delta m = -0.45 \pm 0.2$ and $-0.34 \pm 0.2$
for early and late type galaxies.

\indent
\subsubsection{Selection of galaxies from our SDSS sample} \label{selection}
\indent

We apply the following selection criteria to our 1547 SDSS galaxies
with $v \leq 8000$ km s$^{-1}$.
The limiting magnitude is set at $g'=17.65$, which was
adopted in the luminosity function study by Blanton et al.(2001).
This corresponds to the limit in the absolute magnitude of
$M_{g'} = 2.65-5 $log$_{10}v$.
The limiting magnitude of the ORS is $M_B=-14.0$,
which corresponds to $M_{g'}=-14.45$ and $-14.34$ for early and late type
SDSS galaxies.
Accordingly we also exclude galaxies fainter than this limiting magnitude.
The absolute $g'$ band magnitude
plotted as a function of the recession velocity is shown in Figure~\ref{fig5}.
The sudden change in the density of galaxies which occurs
at around 5000 km s$^{-1}$ reflects a real large scale structure.
Among the 1547 galaxies we find a tiny fraction ($\sim$ 3\%)
of galaxies with $g' \gg 17.65$, which may be assigned unrealistically
faint absolute magnitudes because of some problem in deblending.
There is concern that these galaxies are actually more luminous than
$g' = 17.65$.
They are not included either in the sample or in Figure~\ref{fig5}
because of the above cut in the absolute magnitude.
However, rejection or inclusion of this tiny fraction of erroneous data
does not affect our results.
Our final SDSS sample contains 1078 galaxies.
Figure~\ref{fig6} shows the distribution of galaxies in our SDSS 
sample and ORS galaxies in the EDR region in the equatorial coordinate.
The numbers of observed galaxies in our SDSS and ORS sample within each region
are tabulated in Table~\ref{num}.
Note that the whole region covered by the SDSS sample
is located in the area covered by UGC-m (See Figure~\ref{fig1}).


\indent
\subsubsection{Selection function} \label{sf}
\indent

The selection function of the SDSS can be computed by
using the luminosity function (Blanton et al. 2001).
The luminosity function can be expressed by the Schechter function as
\begin{equation}
\Phi(M) \propto 10^{-0.4(M-M_{*})(\alpha + 1)} \times \exp \left[10^{-0.4(M-M_{*})}\right] \quad,
\end{equation}
where M is the absolute magnitude.
The parameters for $g'$ band determined 
by Blanton et al.(2001) are $M_{*} = -20.04 \pm 0.04$ mag,
$\alpha = -1.26 \pm 0.05$
for $\Omega_{m} = 0.3$ , $\Omega_{\Lambda} = 0.7$ cosmology.
When we restrict our analysis
to galaxies with absolute magnitudes $M < M_{0}$,
the selection function is given by
\begin{equation}
\phi_{\rm SDSS}(r) = \frac{\int_{-\infty}^{M_{\rm max}(r)} \Phi(M) {\mathrm d} M}{\int_{-\infty}^{M_{0}} \Phi(M) {\mathrm d} M},
\label{selecsdss}
\end{equation}
where $M_{\rm max}(r) =  $min$(M_{0},M_{\rm lim}(r))$
and $M_{\rm lim}(r)$ is the limiting magnitude
for a galaxy at distance $r$ to be included into the sample.
We set $M_{0}$ to be $-14.45$ and $-14.34$ for early and late type galaxies
which are the absolute magnitudes in $g'$ band corresponding to $M_B = -14$.
The SDSS selection function obtained in this way is
shown in Figure~\ref{fig3}.

\indent
\section{GALACTIC ABSORPTION} \label{obsc}
\indent

Galactic absorption introduces a nonuniform,
direction-dependent selection effect.
The effective magnitude cutoff limit for a galaxy lying in the direction
($l, b$) is given by $m_{\rm lim}=m_{\rm lim,obs}-A_{\lambda}(l, b)$,
where $A_{\lambda}$ is the absorption at a given band
and ($l, b$) is the Galactic coordinate.
The selection function with the absorption taken into account, is now given
for each of the surveys as
\begin{equation}
\phi_{\rm obs}(r,l,b) = \frac{\int_{-\infty}^{M_{\rm max}(r,l,b)} \Phi(M) {\mathrm d} M}{\int_{-\infty}^{M_{0}} \Phi(M) {\mathrm d} M},
\end{equation}
where 
\begin{eqnarray}
M_{\rm max}(r,l,b) &=& M_{\rm max}(r)+A_{\lambda}(l, b) \\
&=& M_{\rm max}(r \times 10^{0.2 A_{\lambda}(l, b)}).
\end{eqnarray}
Note that $M_{\rm max}(r)$ is related to the flux limit $f_{\rm lim}$ as
$M_{\rm max}(r) = {\rm const} -2.5 \,{\rm log}_{10}\, 4\pi r^2 f_{\rm lim}$.
Thus we can obtain the selection function $\phi_{\rm obs}(r,l,b)$
in which the absorption is taken into account from the one $\phi(r)$
which incorporates no absorption as
\begin{equation}
\phi_{\rm obs}(r,l,b)=\phi(r \times 10^{0.2 A_{\lambda}(l, b)}).
\end{equation}

The ORS gives the absorption $A_B$ to individual galaxy based on
Burstein and Heiles (1982) while the SDSS gives the absorption $A_{g'}$
based on Schlegel et al.(1998).
As shown in Fig~\ref{fig7},
a comparison between the ORS $A_B$ and the SDSS $A_{g'}$ reveals a
systematic difference in the region of low absorption
where $A_B \sim 0.0$ while $A_{g'} \sim 0.1$.
Accordingly we make two computations,
one with $A_B$ given in ORS and the other with $A_B$ computed
from Schlegel et al.(1998) as $A_B=4.1 E(B-V)$,
in the comparison of the galaxy number counts.
The absorption for the SDSS galaxies are computed by $A_{g'}=3.793 E(B-V)$
(Stoughton et al. 2002).


The ORS is based on the isophotal magnitude as mentioned above.
Absorption dims the isophotal magnitude
because not only every point in the galaxy
become dimmer, but the radius of the isophote shrinks.
In Santiago et al.(1996), considering these effects, observed magnitude
is given by $m_{B,\rm{obs}}=m_B+\gamma_m A_B(l, b)$,
where $\gamma_m$ is a free parameter of the selection function
to be determined from the data themselves.
We expect $\gamma_m > 1.0$ in the isophotal magnitude,
the exact value depending on the two-dimensional light profile of the object.
(If $m_{B,\rm{obs}}$ is the total magnitude, $\gamma_m = 1.0$.)
The parameters $\gamma_m$ derived in Santiago et al.(1996) are
1.0 for ESO-m, 0.56 for UGC-m.
However Santiago et al.(1996) noted that errors for $\gamma_m$
are large and these values are consistent with $\gamma_m = 1$.
Since we can not adopt a value of $\gamma_m$ greater than unity
appropriately, we set $\gamma_m=1$ for the ORS throughout the paper.
The SDSS adopt Petrosian magnitude as mentioned in Section~\ref{sdss}.
In this case, the Petrosian radius of the object does not shrink.
Thus we also set $\gamma_m=1$ for the SDSS throughout the paper.

\indent
\section{CALIBRATION OF THE ORS SELECTION FUNCTION} \label{calibration}
\indent

In this section, we compare the galaxy number counts derived from our SDSS
sample (1078 galaxies) with those from our ORS sample (130 galaxies).
We start by counting the number of the observed galaxies in the SDSS
and in the ORS, which lie in each of the regions covered by the SDSS,
in velocity bins of width 1000 km s$^{-1}$. 
Then, we correct this number for the selection effect
using respective selection functions shown in Figure~\ref{fig3}.
The galactic absorption for the ORS is computed
following Schlegel et al.(1998).
The results are summarized in Table~\ref{num} and shown
in the upper panels of Figure~\ref{fig8}.
The error bars are defined so as to represent a square root of
the number of galaxies in each velocity bin.
Since observed galaxies in the SDSS are much more than in the ORS,
we show errorbars only for the ORS.
We find that number counts of the ORS are systematically
larger outside 5000 km s$^{-1}$.


Cosmic variance is always a concern in this kind of analysis.
For this reason, we counted the galaxies separately in each of
the three regions.
As shown in Table~\ref{num} and Figure~\ref{fig8},
we found the overestimate of the ORS in all the three regions.
Although we can not draw strong conclusion as for the significance
of this overestimate considering the current small area coverage of the SDSS,
accumulation of the data may allow us to do more
extensive investigation in the near future.
For the moment, we regard the systematic difference as significant
because we have carefully counted galaxies in the same magnitude range
($M_{B} < -14$) and the SDSS spectroscopic catalog has a very high
completeness except for the bright end as shown in Strauss et al.(2002).

We revise the selection function of UGC-m by changing the
parameter $b$, which is sensitive at large distances,
so that the ORS galaxy counts coincide with the SDSS counts.
The effect of changing $b$ is clearly visible in Figure~\ref{fig9}, where
the galaxy number counts of the ORS in ALL region
are shown for various choices of the parameter $b$.
The revised values of $b$ for regions 1, 2, 3, and ALL region
for the adopted values of the magnitude offset of $g'-B_J=0.34$ mag
(early type) and $-0.45$ mag (late type)
are summarized in the top line of Table~\ref{b}.
Absorption averaged over ALL region is also given in this table.
Comparison of the galaxy number counts using the revised
selection function are shown in the lower panels of Figure~\ref{fig8}.
We show the revised selection function for ALL region in
Figure~\ref{fig10}, together with that by Santiago et al.(1996).


We examine the effect of uncertainty of the offset.
The values of $b$ for $\Delta m=-0.1 \pm 0.2$ mag are also shown
in Table~\ref{b}.
The $b$ values changes by $\sim \pm 0.3$ for each region.
Furthermore, we also tabulate results derived by using
the original absorption given in the ORS catalog in Table~\ref{b}.
Replacing the original ORS absorption with that based on Schlegel et al.(1998)
changes the $b$ values by $\sim 0.9$, which is larger than the change
caused by the different offset values.
In all cases, the values of $b$ become smaller than
that derived by Santiago et al.(1996) ($b_{UGC}=9.38$).


\indent
\section{SPATIAL STRUCTURE OF THE LOCAL SUPERCLUSTER} \label{model}
\indent

It is long known that there is an excess of galaxies in the vicinity 
of the north galactic pole, which is now referred to as the LSC
(e.g., de Vaucouleurs 1956; Yahil et al. 1980; Tully 1982).
Tully (1982) illustrated the three-dimensional distribution
of galaxies in the LSC and described its spatial structure as consisting
of two components, one is a flattened disk and the other is a
roughly spherical halo.

Only luminous galaxies are often counted as sources of UHECRs,
and faint galaxies are neglected.
However, the boundary between luminous and faint galaxies is not
established at all in the context of acceleration scenarios.
If luminous galaxies in the LSC are distributed
differently from fainter ones, the distribution of arrival directions
of UHECRs may change with the limiting absolute magnitude
of the sample considered.
Thus we determine the spatial structure
of the LSC using galaxies of different absolute magnitudes.
In Figure~\ref{fig11}, we show distribution of galaxies
more luminous than $M_{B} = -14,-16,-18$ in the LSC
in the super-galactic coordinate.
The two components, the flattened disk and the roughly spherical halo,
are clearly visible.
Here we restrict our analysis to galaxies within 2000 km s$^{-1}$
from the Virgo cluster (SGX=SGZ=0 km s$^{-1}$, SGY=1000 km s$^{-1}$).


We assume that the number density of the halo is uniform while
that of the disk is exponential both in the radial and the z-directions.
Thus, the number density of galaxies of our model
is expressed as
\begin{equation}
n(\rho,z) \propto \left( {\exp} \left(-\frac{\rho}{\alpha}\right) {\exp} \left(-\frac{|z|}{\beta}\right)+n_{1} \right) ,
\label{modeleq}
\end{equation}
where $\rho,z$ are the cylindrical coordinates 
in which the Virgo Cluster is located at the origin ($\rho=0,z=0$)
and the plane of the disk component coincides with the $z=0$ plane.
The parameter $n_{1}$ in Eq.(~\ref{modeleq}) represents the ratio of 
the halo component to the disk component.
The parameters of $\alpha$ and $\beta$ are the radial and vertical
scale lengths of the number density of galaxies
in the disk component, respectively.

The probability $f(\overrightarrow{r_{i}})$ 
that a galaxy is observed at position $\overrightarrow{r_{i}}$
is proportional to the number density multiplied by the selection function
at $\overrightarrow{r_{i}}$ and
divided by the integral of these values over the region 
which we consider, that is, the region 
within 2000 km s$^{-1}$ (20 $h^{-1}$ Mpc) from the Virgo cluster.
Thus we can write
\begin{equation}
f(\rho,z)=n(\rho,z)\phi_{ORS}(\rho,z)/\int n(\overrightarrow{r}) \phi_{ORS}(\overrightarrow{r})  {\mathrm d}^3 r.
\label{prob}
\end{equation} 
This probability function is independent of the normalization 
of the number density.
For computational convenience we define a likelihood function
to be the logarithm of the product of Eq.(~\ref{prob})
over all galaxies in the sample and in the region considered:
\begin{equation}
\Lambda = \sum _{i} {\ln} \left( \frac{n(\overrightarrow{r_{i}}) \phi_{ORS}(\overrightarrow{r_{i}})}{\int n(\overrightarrow{r}) \phi_{ORS}(\overrightarrow{r})  {\mathrm d}^3 r} \right).
\label{tprob}
\end{equation}
We determine the most probable values of $\alpha,\beta$ and $n_{1}$ 
by maximizing Eq.(~\ref{tprob}) with respect to these parameters
for each of the three samples with $M_B < -14,-16$ and $-18$.
In the calculation of Eq.(~\ref{tprob}), we take account of the
absorption from Schlegel et al.(1998).
The ORS excluded the region of $|b|<20^{\circ}$, and in addition,
the region at higher latitude in which the B-band extinction
according to Burstein and Heiles (1982) is
greater than $A_B=0.7$ mag (this excludes only $0.18$ sr).
However, a tiny fraction ($\sim 4 \%$) of our ORS sample lie at the position
where the absorption from Schlegel et al.(1998) is larger than $A_B=0.7$ mag.
We exclude those galaxies in the calculation of Eq.(~\ref{tprob}).

In order to determine the most probable values, we need to obtain
the revised selection function for ESO-m as well as for UGC-m.
The revised $b$ value for ESO-m is derived as follows.
We compute the ratio of the value of the original UGC-m selection
function to that of the revised selection function at 8,000 km s$^{-1}$,
and then, calculate the revised $b$ value for the ESO-m selection
function so that the above ratio takes the same value
as for the UGC-m selection function.
Because we want to determine the most probable values for galaxies with
different absolute luminosities, we compute the selection function
corresponding to each limiting magnitude as follows.
First, we obtain the luminosity function of the ORS by differentiating
Eq.(1) with respect to $r$.
Note that the derivative of the selection function gives the
luminosity function as seen, for example, Eq.(\ref{selecsdss}).
With this luminosity function, we calculate the ORS selection
function by the equation similar to Eq.(\ref{selecsdss})
with various values of $M_0$.
We choose $M_0 = -16$ and $-18$ as well as the original choice of $M_0=-14$.
In Figure~\ref{fig12}, the original and the revised selection
functions of UGC-m and ESO-m samples are shown for $M_B<-14,
-16$ and $-18$. It is noted that for $M_B<-18$ both UGC-m and ESO-m 
samples are complete to $\sim$ 3000 km s$^{-1}$ (see also Figure~\ref{fig2}).


We show the values of the likelihood function
with respect to $\alpha, \beta$ and $n_{1}$ in Figures~\ref{fig13}.
The most probable values of the model parameters are summarized
in Table~\ref{disk}.
It is found that the revision of the selection function
does not affect the results in this analysis.
Most probable values of $\beta$ are almost the same for
all the luminosity ranges, while $\alpha$ and $n_{1}$ are larger
for $M_B < -18$ than for $M_B < -14$ and $-16$.
In other words, if we do not see dwarf galaxies ($M_B \ge -16$),
the disk looks more extended outward and halo looks denser.
These properties are also seen in Figure~\ref{fig11}.


Let us calculate a quantitative value of the
local overdensity due to the disk component in the LSC.
We define the local overdensity $\delta$
as the ratio of number density of galaxies averaged over the
disk component to that of the halo component, which is assumed to
be uniform in this paper.
(We tacitly assume all the galaxies to have same mass.)
We define the volume of the disk component over which the number density
is averaged by
\begin{equation}
\frac{\rho}{\alpha (h^{-1}\mbox{Mpc})}+\frac{z}{\beta (h^{-1} \mbox{Mpc})} \leq -ln(n_{1}).
\end{equation}
The volume corresponds to the region where the first term
in Eq.(~\ref{modeleq}) which expresses the disk is larger than
the second term which expresses the halo.
The local overdensities for different luminosity ranges
are also given in Table~\ref{disk}.
Baker et al.(1998) calculated the density field contours
on several radial shells with the Gaussian smoothing,
and obtained local overdensity $\sim 10$ for the Virgo cluster
using the ORS galaxies.
Strictly speaking, local overdensity derived by Baker et al.(1998) is not
defined in the same manner as ours.
However, the amount of the local overdensity derived by our number-density
model is consistent with that obtained by Baker et al.(1998).


\indent
\section{SUMMARY AND DISCUSSION} \label{summary}
\indent

In this paper, we investigate the uncertainty of the selection function
of the ORS, which we used as sources of UHECRs,
by comparing the galaxy number counts of ORS galaxies with those of SDSS EDR
galaxies as a function of recession velocity.
We carefully count galaxies from the two data samples so that these galaxies
are in the same absolute luminosity range.
In all the regions covered by the SDSS EDR, we find
that the ORS predicts systematically more galaxies than the SDSS 
by a factor of $\sim$ 2 outside 5000 km s$^{-1}$.
We revise the selection function of the ORS so that the predictions of
the two surveys become consistent.
It is also found that replacing the original absorption given in the ORS
catalog with that computed from Schlegel et al.(1998) changes the
$b$ parameter of the ORS selection function by more than 10 $\%$.
The revision should be taken into account in the future analysis
of the source number density of UHECRs based on the ORS.

Using this revised selection function of the ORS,
we determine the spatial structure of the LSC with a model
consisting of a uniform spherical halo and an exponential disk.
The model parameters are derived
for galaxies in three ranges of different absolute luminosities,
$M_B < -14$, $M_B < -16$ and $M_B < -18$.
The most probable values of the scale length of the disk in $z$
direction, $\beta$, are almost the same for the three luminosity ranges
while the radial scale length of the disk, $\alpha$, and the density
of the halo, $n_1$, are larger for $M_B < -18$ than for $M_B < -14$ and $-16$.
This implies that giant galaxies $(M_B < -18)$ in the disk are
more extended radially than dwarf galaxies $(M_B > -16)$ and
that giant galaxies show a smaller local overdensity than dwarf galaxies.

Let us consider the effect of this difference of the distribution
of galaxies in the LSC on the distribution of UHECR arrival directions.
Observationally, it is reported that there is no significant
large-scale anisotropy in the arrival direction distribution of UHECRs
(Takeda et al. 1999).
If all the galaxies, both giants and dwarfs, are sources of UHECRs,
arrival directions could be more anisotropic than the case that
only luminous giant galaxies are responsible for UHECRs.
The observed lack of anisotropy may favor the latter case.
Provided that UHECRs are generated in star forming regions,
their sources are expected to trace the distribution of late-type galaxies
rather than early-type galaxies.
It is known that there is strong spatial
segregation of the Hubble types in the Virgo cluster (Binggeli et al. 1987).
Early-type galaxies are more concentrated towards the cluster center
than late types.
Distribution of galaxies of different morphological types in the LSC
is also an important factor to be taken into account in the analysis of
arrival direction of UHECRs.

We restricted ourselves in this paper to the study of galaxies
as the sources of UHECRs.
The revision of the ORS selection function made in this study
is found to be insignificant in the determination of the global
structure of the LSC.
The revised selection function obtained in this study will, however, be
useful to other studies such as peculiar velocity and correlation function.
The SDSS will produce imaging and spectroscopic surveys
in five bands ($u',g',r',i',z'$)
over $\pi$ steragians in the northern Galactic cap,
and will collect spectra of approximately 1,000,000 galaxies
(down to $r'_{lim} \sim 17.65$).
We have used in this paper the SDSS EDR which covers
only less than $1\%$ of all sky.
As data obtained by the SDSS accumulate,
we will be able to revise the selection function of several
surveys with better accuracy, and also to determine the spatial structure
of the LSC using only the SDSS galaxies.

\section*{Acknowledgments}
The authors are grateful to Dr. K. Shimasaku and SDSS
group in University of Tokyo for useful discussions.
This research was supported in part by Giants-in-Aid for Scientific
Research provided by the Ministry of Education, Science and Culture
of Japan through Research Grant No.S14102004.

\section*{References}
\re
Baker J.E., Davis M., Strauss M.A., Lahav O., Santiago B.X. 1998, \apj, 508, 6
\re
Binggeli B., Tammann G.A., Sandage A. 1987, \aj, 94, 251
\re
Blanton M., Blasi P., Olinto A.V. 2001, Astropart. Phys., 15, 275
\re
Blanton M.R., et al. 2001, \aj, 121, 2358
\re
Burstein D., and Heiles C. 1982, \aj, 87, 1165
\re
Couch, W.J., and Newell, E.B. 1980, PASP, 92, 746
\re
de Vaucouleurs, G. 1956, Vistas in Astronomy, 2, 1584
\re
Fukugita M., Shimasaku K., Ichikawa T. 1995, PASP, 107, 945
\re
Greisen K. 1966, Phys. Rev. Lett., 16, 748
\re
Ide Y., Nagataki S., Tsubaki S., Yoshiguchi H., Sato K.
2001, PASJ, 53, 1153
\re
Lemoine, M., Sigl, G., and Biermann, P. 1999, astro-ph/9903124
\re
Lupton R. H., Gunn J. E., and Szalay A. S. 1999, \aj, 118, 1406
\re
Petrosian V., 1976, \apj, 209, L1
\re
Santiago B.X., Strauss M.A., Lahav O., Davis M., Dressler A., Huchra J.P.
1995, \apj, 446, 457
\re
Santiago B.X., Strauss M.A., Lahav O., Davis M., Dressler A., Huchra J.P.
1996, \apj, 461, 38
\re
Schlegel D., Finkbeiner D., and Davis M. 1998, \apj, 500, 525
\re
Sigl, G., Lemoine, M., and Biermann, P. 1999, Astropart. Phys., 10, 141
\re
Smialkowski, A., Giller, M., and Michalak, W. 2002, astro-ph/0203337
\re
Stoughton C., et al. 2002, \aj, 123, 485
\re
Strateva I., et al. 2001, \aj, 122, 1861
\re
Strauss M. A., et al. 2002, \aj, submitted
\re
Takeda M., et al. 1999, \apj, 522, 225
\re
Tully R.B. 1982, \apj, 257, 389
\re
Yahil, A., Sandage A., and Tammann G.A. 1980, Astrophys. J., 242, 448
\re
Yasuda N., Fukugita M., Narayanan V. et al. 2001, \aj, 122, 1104
\re
Yoshiguchi H., Nagataki S., Tsubaki S., Sato K.
2002, astro-ph/0210132
\re
Zatsepin G.T., Kuz'min V.A. 1966, J. Exp. Theor. Phys. Lett., 4, 78
\re

\clearpage
\begin{table*}
\begin{center}
\begin{tabular}{r|cc|cc|cc|cc|cc|cc}    \hline
    & \multicolumn{4}{|c|}{Region 1} & \multicolumn{4}{|c|}{Region 2} &
\multicolumn{4}{|c}{Region 3} \\ \hline
\multicolumn{1}{c|}{$v$} &
\multicolumn{2}{|c|}{ORS} & \multicolumn{2}{|c|}{SDSS} &
\multicolumn{2}{|c|}{ORS} & \multicolumn{2}{|c|}{SDSS} &
\multicolumn{2}{|c|}{ORS} & \multicolumn{2}{|c}{SDSS}  \\ \hline
\multicolumn{1}{c|}{(km s$^{-1}$)} &
$N_{\rm {obs}}$ & $N_{\rm {cor}}$ & 
$N_{\rm {obs}}$ & $N_{\rm {cor}}$ & 
$N_{\rm {obs}}$ & $N_{\rm {cor}}$ & 
$N_{\rm {obs}}$ & $N_{\rm {cor}}$ & 
$N_{\rm {obs}}$ & $N_{\rm {cor}}$ & 
$N_{\rm {obs}}$ & $N_{\rm {cor}}$ \\ \hline\hline
0-1000      & 0 & 0.0   & 0   & 0.0   & 2   & 3.3   & 1   & 1.0   &
              0 & 0.0   & 1   & 1.0   \\
1000-2000   & 0 & 0.0   & 0   & 0.0   & 23  & 62.6  & 46  & 46.0  &
              8 & 20.5  & 14  & 14.0  \\
2000-3000   & 2 & 11.5  & 2   & 2.3   & 6   & 26.3  & 16  & 16.5  &
              4 & 20.9  & 26  & 28.4  \\
3000-4000   & 8 & 59.7  & 30  & 39.4  & 5   & 46.1  & 23  & 31.9  &
              1 & 7.9   & 10  & 14.0  \\
4000-5000   & 2 & 29.9  & 10  & 17.6  & 3   & 45.3  & 18  & 30.5  &
              6 & 81.9  & 41  & 68.4  \\
5000-6000   & 1 & 18.1  & 15  & 30.3  & 14  & 398.1 & 79  & 161.6 &
             16 & 405.5 & 177 & 353.4 \\
6000-7000   & 2 & 91.6  & 19  & 44.4  & 7   & 360.7 & 124 & 297.5 &
             15 & 944.0 & 153 & 381.5 \\
7000-8000   & 1 & 143.2 & 36  & 103.1 & 1   & 101.7 & 127 & 363.3 &
              3 & 335.2 & 110 & 318.6 \\ \hline
total       & 16& 354.0 & 112 & 237.1 & 61  & 1044.2& 434 & 948.4 &
             53 & 1815.9& 532 &1179.2 \\ \hline
\end{tabular}
\end{center}
\caption{
\normalsize{
Number of galaxies in our ORS and SDSS samples in each region
defined in Figure 4.
$N_{\rm {obs}}$ and $N_{\rm {cor}}$ represent the number of observed galaxies
and that corrected for the selection effect
using the selection functions, respectively.
The magnitude offset is taken as $g'-B_J+\Delta m = -0.45$ and $-0.34$
for early and late type SDSS galaxies.
}
\label{num}}
\end{table*}

\begin{table}
\begin{center}
\begin{tabular}{cc|r@{.}l|r@{.}lc|r@{.}lr@{.}lr@{.}lr@{.}l}    \hline
\multicolumn{2}{c}{offset} & \multicolumn{5}{|c|}{absorption} &
\multicolumn{8}{|c}{region} \\ \hline

\multicolumn{2}{c}{$g'-B_J+\Delta m$} & \multicolumn{2}{|c|}{SDSS} &
\multicolumn{3}{|c|}{ORS} & \multicolumn{2}{|c}{1}
& \multicolumn{2}{c}{2} & \multicolumn{2}{c}{3} 
& \multicolumn{2}{c}{ALL } \\ \hline

early & late & \multicolumn{2}{|c|}{$A_{g'}$} &
\multicolumn{3}{|c|}{$A_B$} &
\multicolumn{8}{|c}{} \\
\hline\hline
$-0.34$ & $-0.45$ & 0&15  & 0&22  & Schlegel & 7&9 &  7&9 & 6&5 & 7&5\\
$-0.14$ & $-0.25$ & 0&15  & 0&22  & Schlegel & 8&2 &  8&3 & 6&8 & 7&8\\
$-0.54$ & $-0.65$ & 0&15  & 0&22  & Schlegel & 7&6 &  7&5 & 6&2 & 7&2\\   
$-0.34$ & $-0.45$ & 0&15  & 0&07  & original & 8&5 &  8&6 & 7&5 & 8&4\\ \hline
\end{tabular}
\end{center}
\caption{
\normalsize{
Parameters $b$ of the ORS selection function for UGC-m revised in this study
according to several choices of the magnitude offset
between the SDSS and ORS and of the absorption effect.
The original value of $b$ derived by Santiago et al.(1996) is 9.38.
The absorptions averaged over all the galaxies in the ALL region
are also tabulated.
Absorption of the SDSS is computed following Schlegel et al.(1998).
}
\label{b}}
\end{table}

\begin{table*}
\begin{center}
\begin{tabular}{ccccr@{.}lr@{.}l}    \hline
Absolute & \multicolumn{1}{c}{$N_{\rm {gal}}$} & 
\multicolumn{1}{c}{$\alpha$} & \multicolumn{1}{c}{$\beta$} & 
\multicolumn{2}{c}{$n_1$} & \multicolumn{2}{c}{local}
 \\
Luminosity & \multicolumn{1}{c}{}
& \multicolumn{2}{c}{$(h^{-1} Mpc)$} & \multicolumn{2}{c}{} 
& \multicolumn{2}{c}{overdensity } \\ 
\hline\hline
$M_{B}<-14$  & 1291 & 360 & 180 & 0&013 & 10&5 \\
$M_{B}<-16$  & 1137 & 380 & 220 & 0&015 & 10&0 \\
$M_{B}<-18$  & 524  & 480 & 180 & 0&023 &  8&7 \\    \hline
\end{tabular}
\end{center}
\caption{
\normalsize{
Parameters of our number-density model of the Local Supercluster calculated
using the ORS sample of different luminosity ranges within 2000 km s$^{-1}$
from the Virgo cluster.
The local overdensity due to the disk component is also shown (see the text).
}
\label{disk}}
\end{table*}

\thispagestyle{empty}
\begin{figure}
\begin{center}
   \FigureFile(80mm,60mm){fig1.eps}
\end{center}
\caption{
Distribution of galaxies in the ESO-m and UGC-m samples used in the
present study ($v \leq$ 8000 km s$^{-1}$, $M_{B}\leq -14.0$, $m_{B} \leq 14.5$,
$|b| > 20^\circ$).
}
\label{fig1}
\end{figure}

\thispagestyle{empty}
\begin{figure}
\begin{center}
   \FigureFile(70mm,50mm){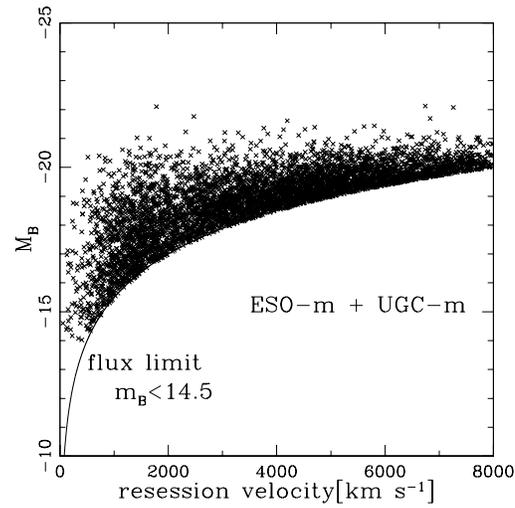}
\end{center}
\caption{
Absolute magnitude plotted as a function of the recession velocity
for galaxies in our ORS sample.
}
\label{fig2}
\end{figure}

\thispagestyle{empty}
\begin{figure}
\begin{center}
   \FigureFile(70mm,50mm){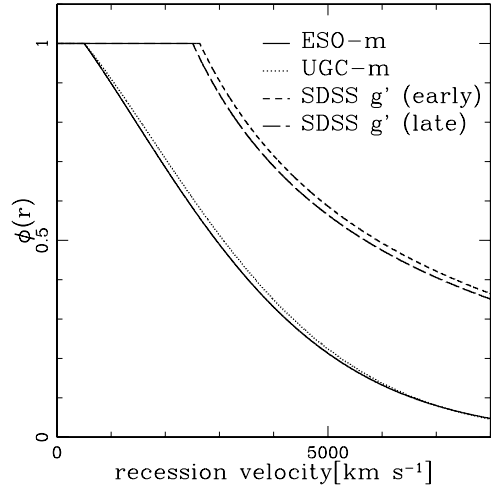}
\end{center}
\caption{
Selection functions of the ESO-m and the UGC-m
given in Santiago et al.(1995) and of the SDSS ($g'$ band).
The SDSS selection function is calculated from the luminosity function
given in Blanton et al.(2001).
}
\label{fig3}
\end{figure}

\thispagestyle{empty}
\begin{figure}
\begin{center}
	\FigureFile(80mm,60mm){fig4.eps}
\end{center}
\caption{
Projection in equatorial coordinate of the regions which were covered by the
spectroscopy of the SDSS EDR.
}
\label{fig4}
\end{figure}

\thispagestyle{empty}
\begin{figure}
\begin{center}
   \FigureFile(60mm,50mm){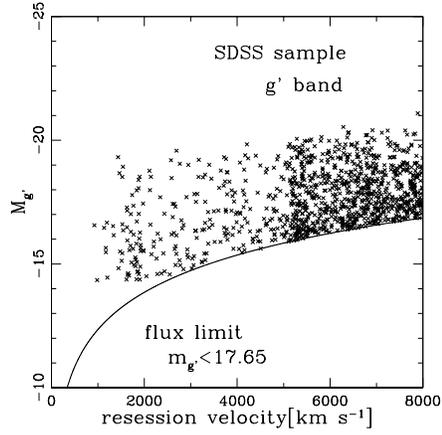}
\end{center}
\caption{
Absolute $g'$ band magnitude plotted as a function of
the recession velocity for galaxies in our SDSS sample.

}
\label{fig5}
\end{figure}

\thispagestyle{empty}
\begin{figure}
\begin{center}
	\FigureFile(60mm,50mm){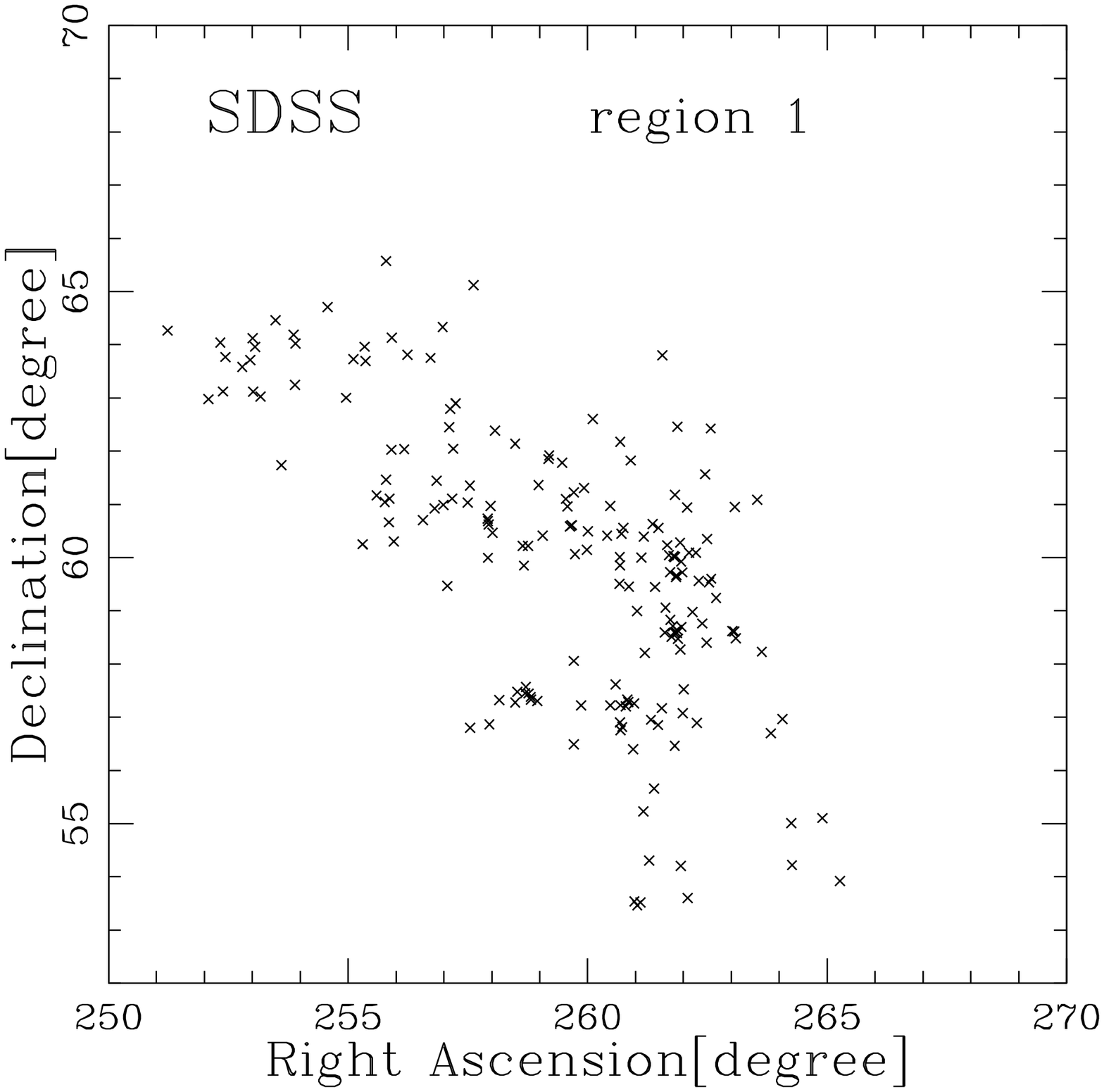}
	\FigureFile(60mm,50mm){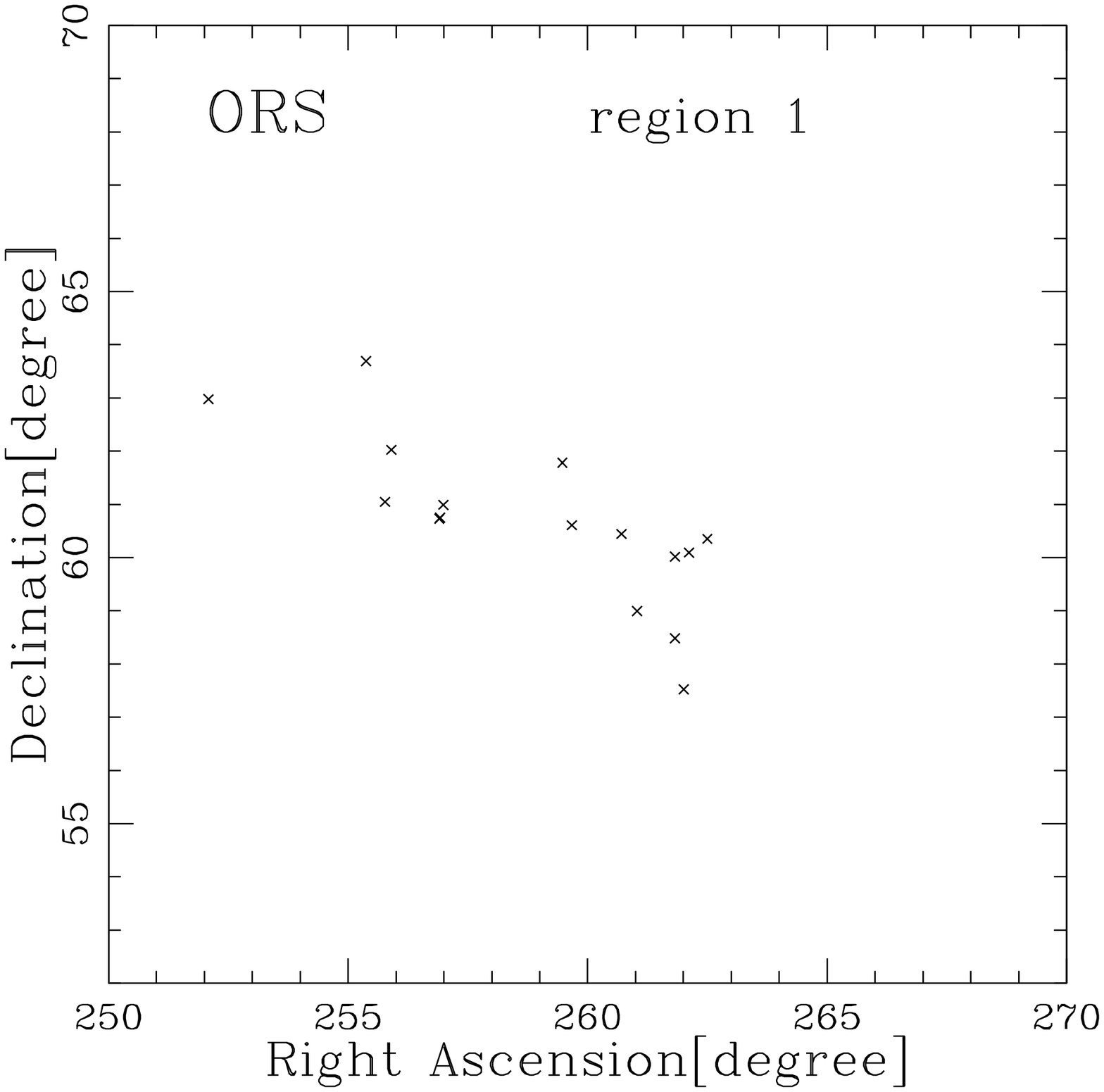}
	\FigureFile(60mm,50mm){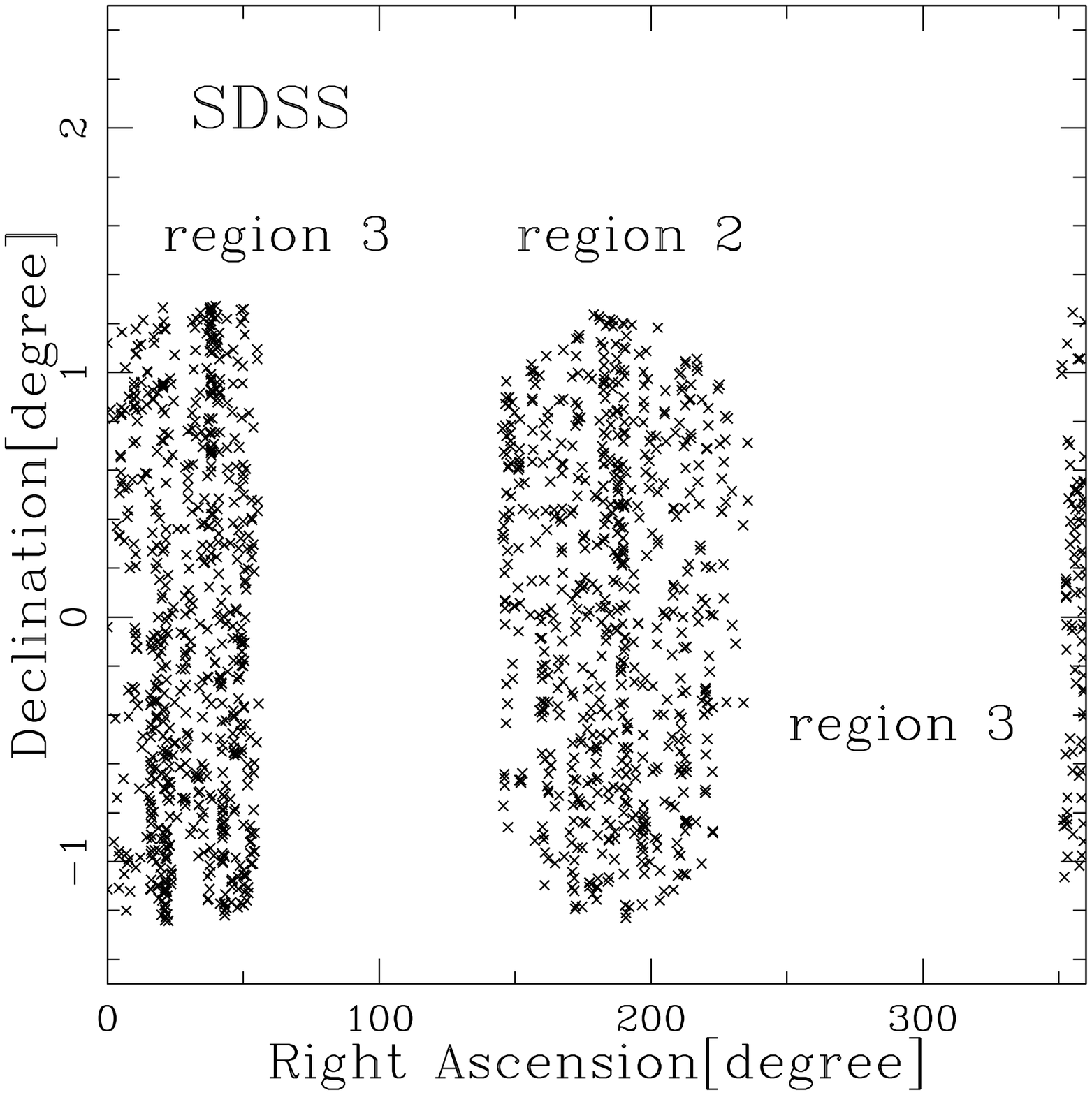}
	\FigureFile(60mm,50mm){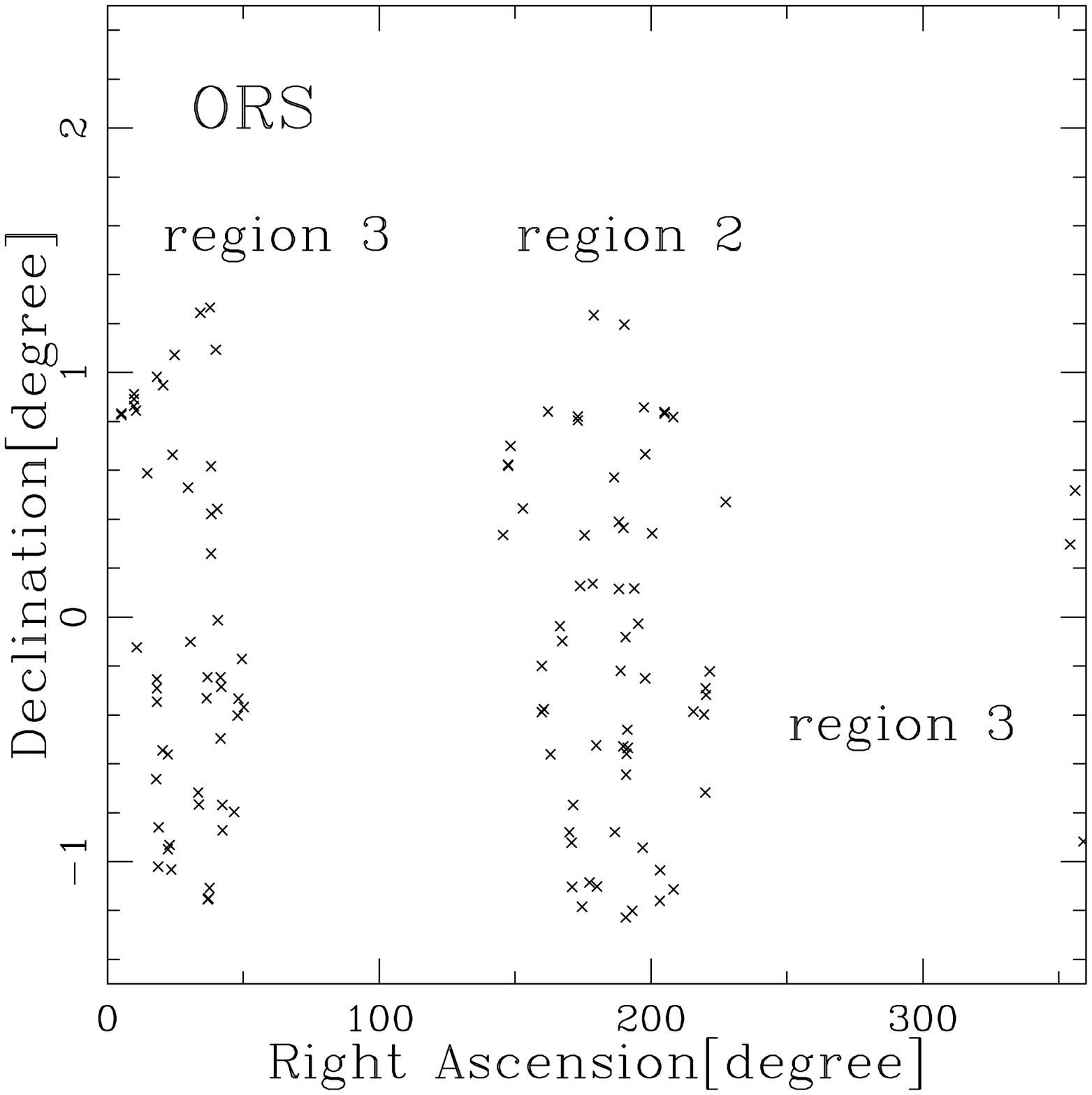}
\end{center}
\caption{
Distribution of 1078 galaxies in our SDSS sample (left panels)
and 130 galaxies in our ORS sample (right panels).
}
\label{fig6}
\end{figure}

\thispagestyle{empty}
\begin{figure}
\begin{center}
	\FigureFile(60mm,50mm){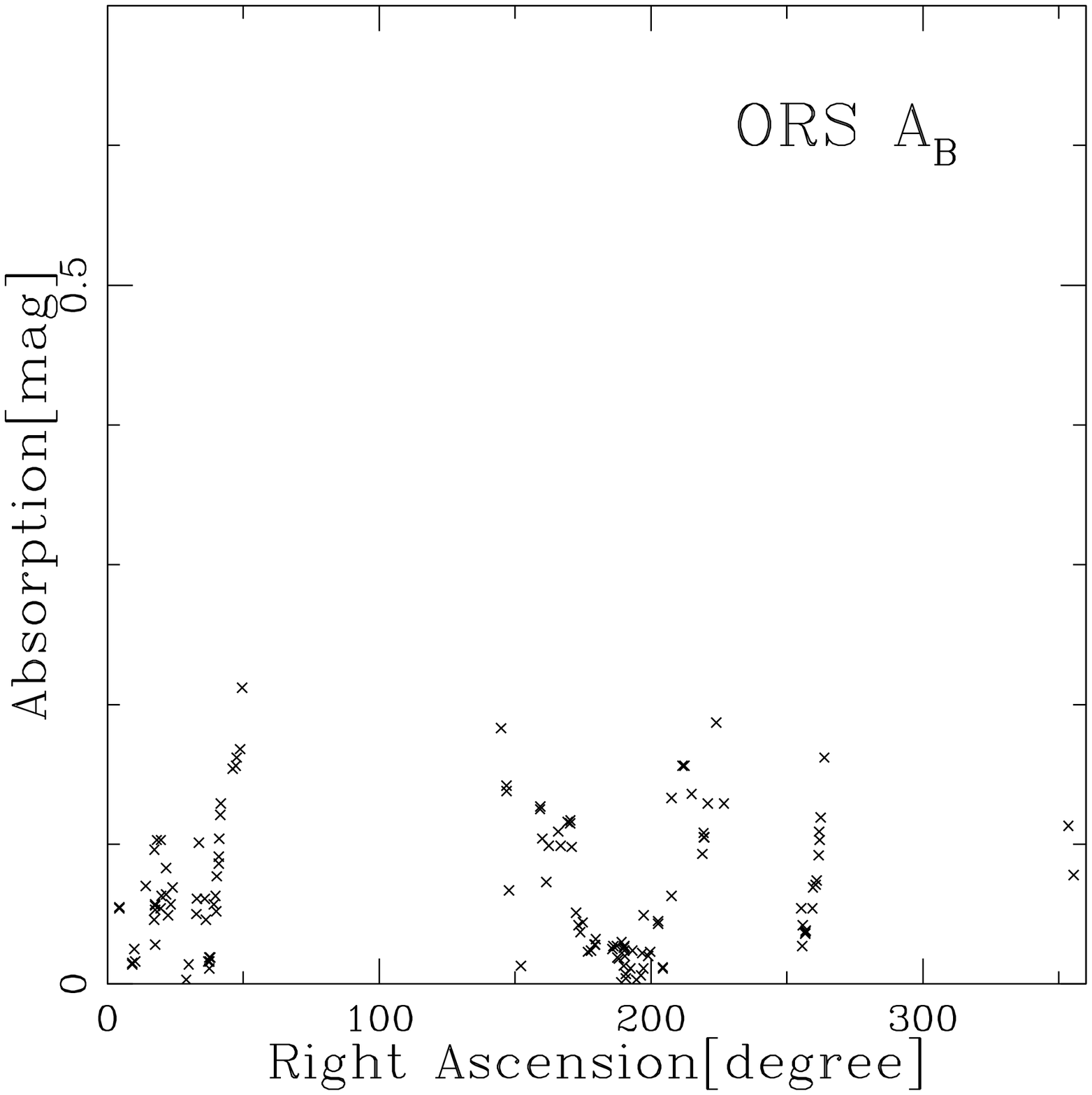}
	\FigureFile(60mm,50mm){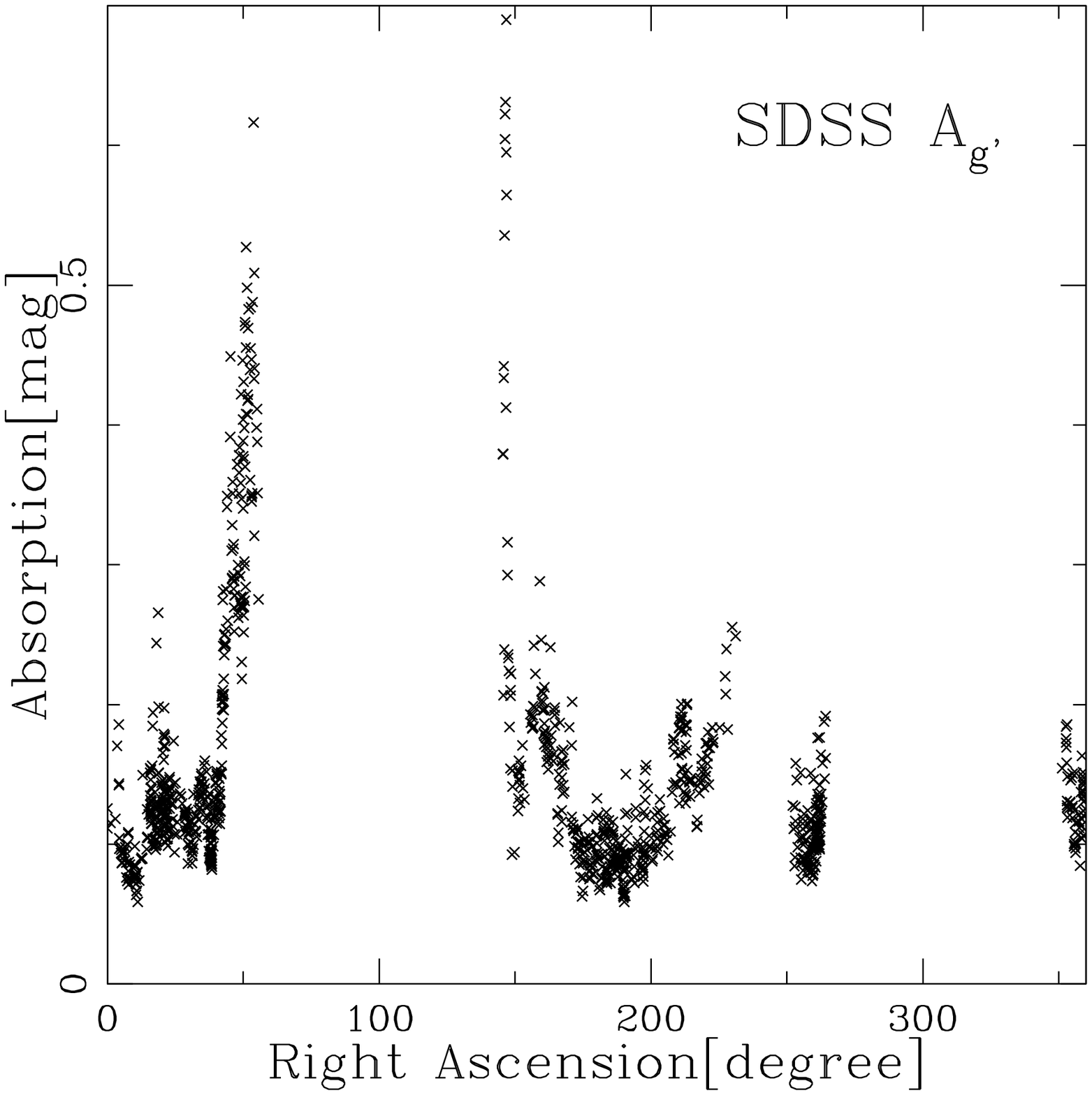}
\end{center}
\caption{
Absorptions given to individual ORS galaxies
in the regions of SDSS EDR (left panel)
and those given to SDSS galaxies (right panel) as a function of
right ascension.
A systematic difference between the two in the region of low
absorption is clearly visible.
}
\label{fig7}
\end{figure}

\thispagestyle{empty}
\begin{figure}
\begin{center}
	\FigureFile(40mm,40mm){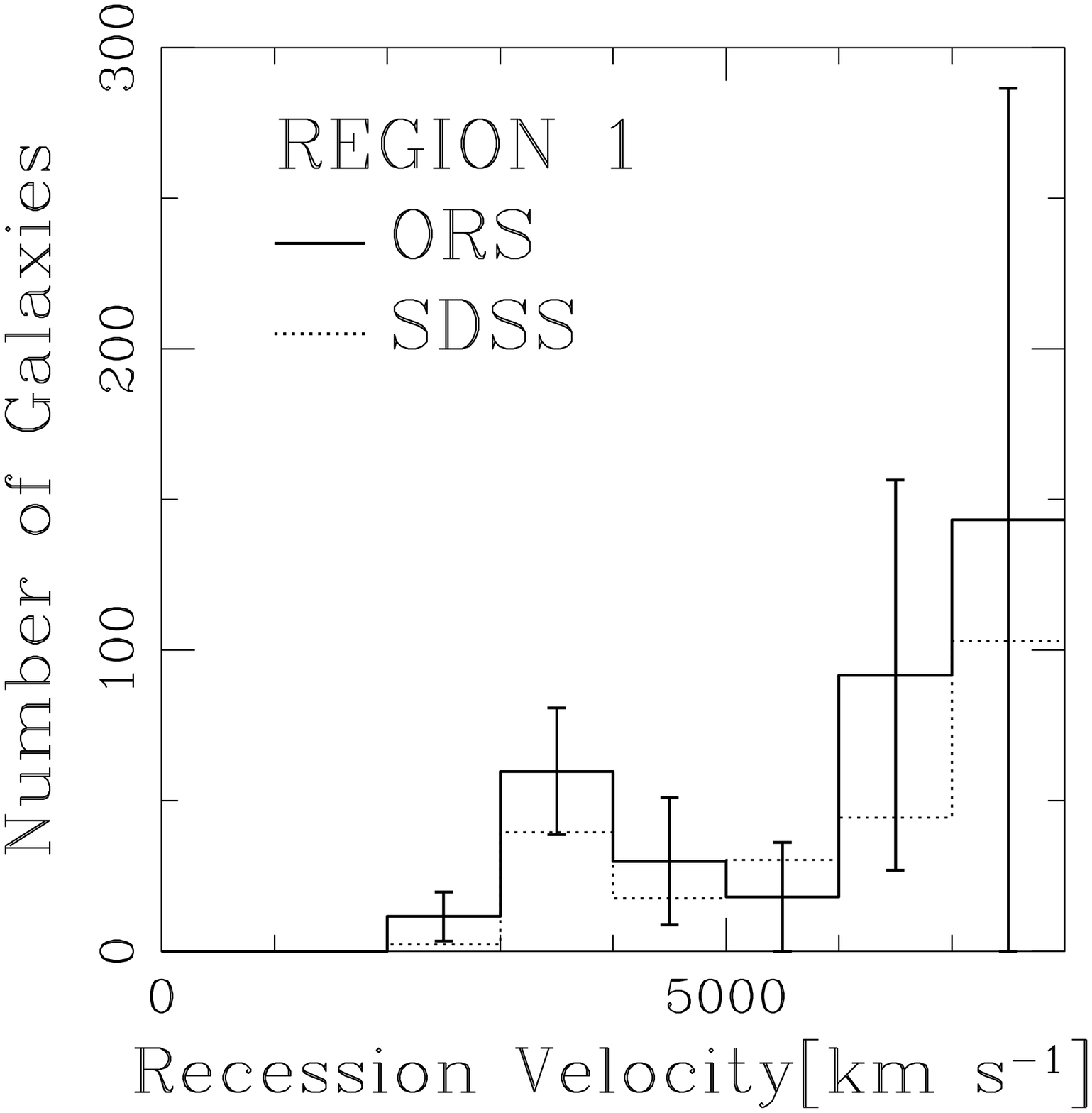}
	\FigureFile(40mm,40mm){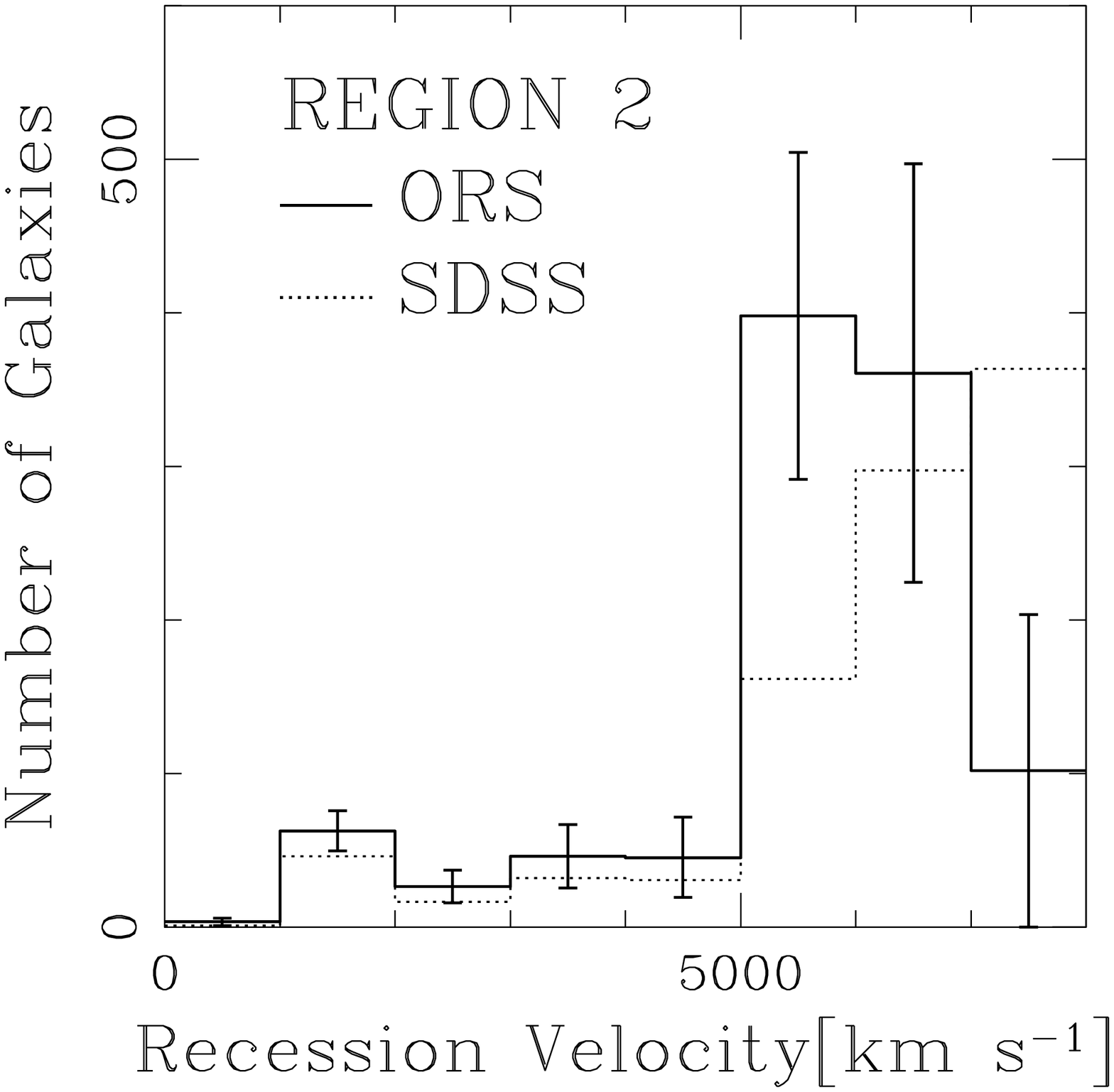}
	\FigureFile(40mm,40mm){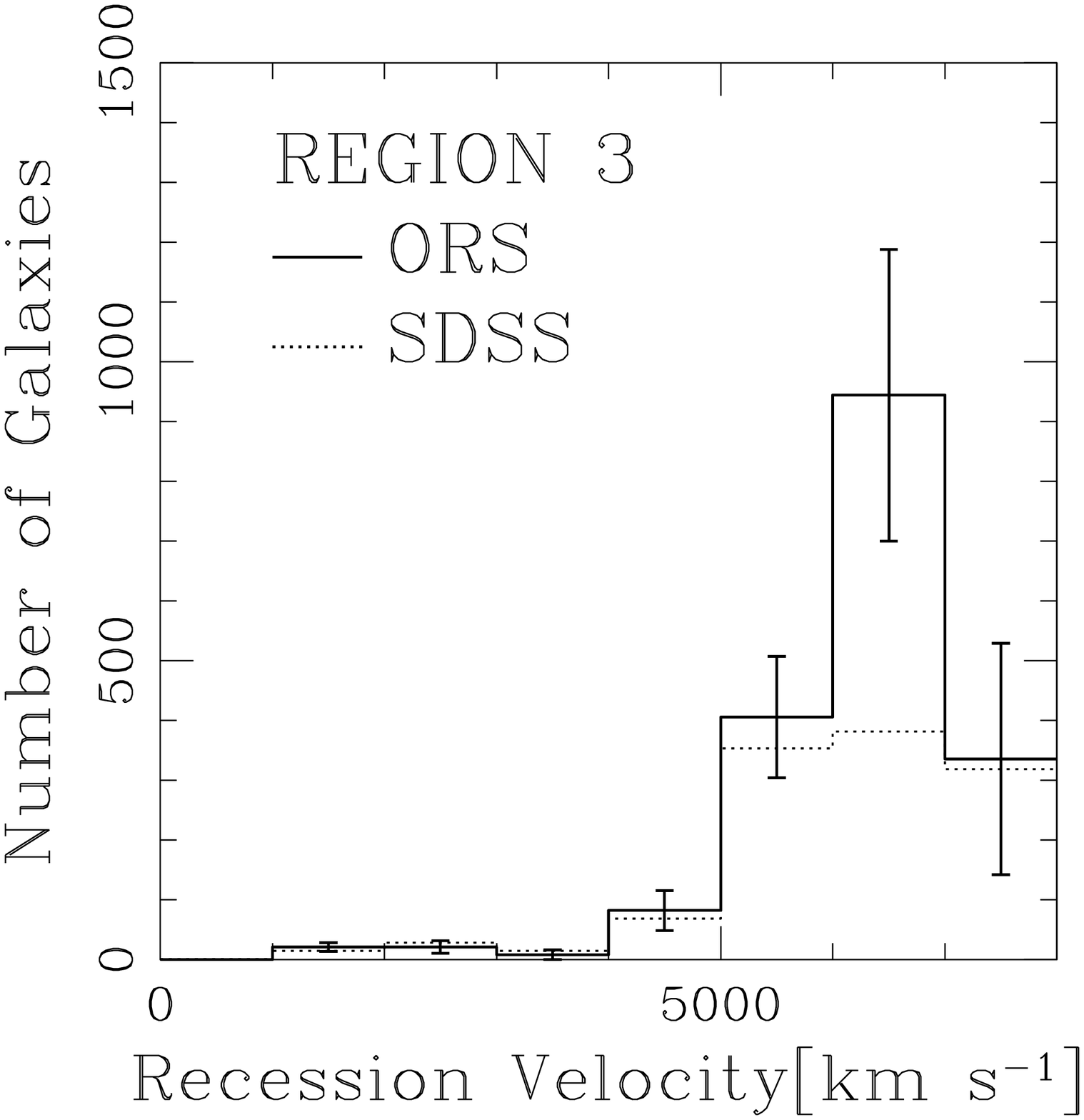}
	\FigureFile(40mm,40mm){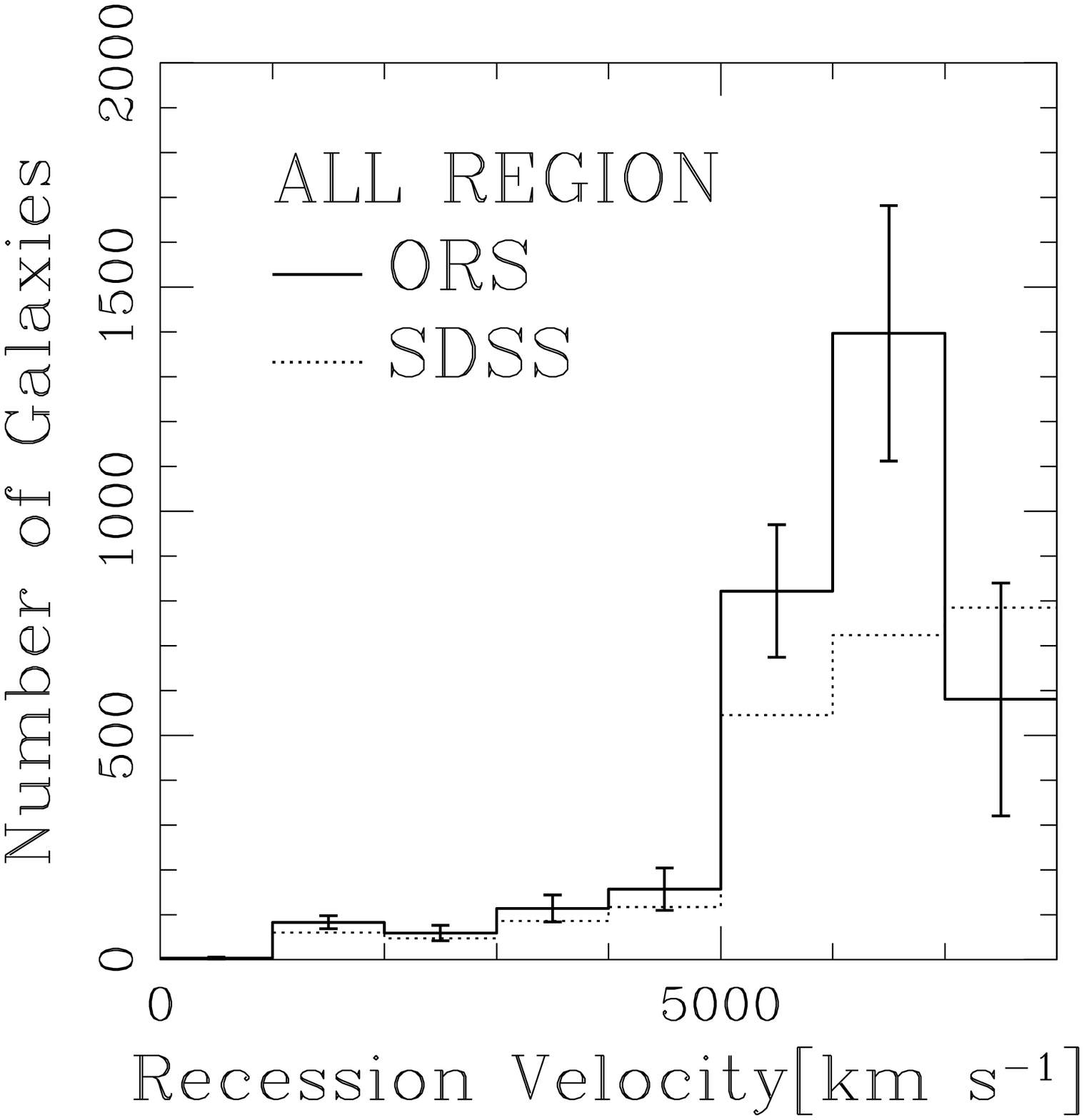}
	\FigureFile(40mm,40mm){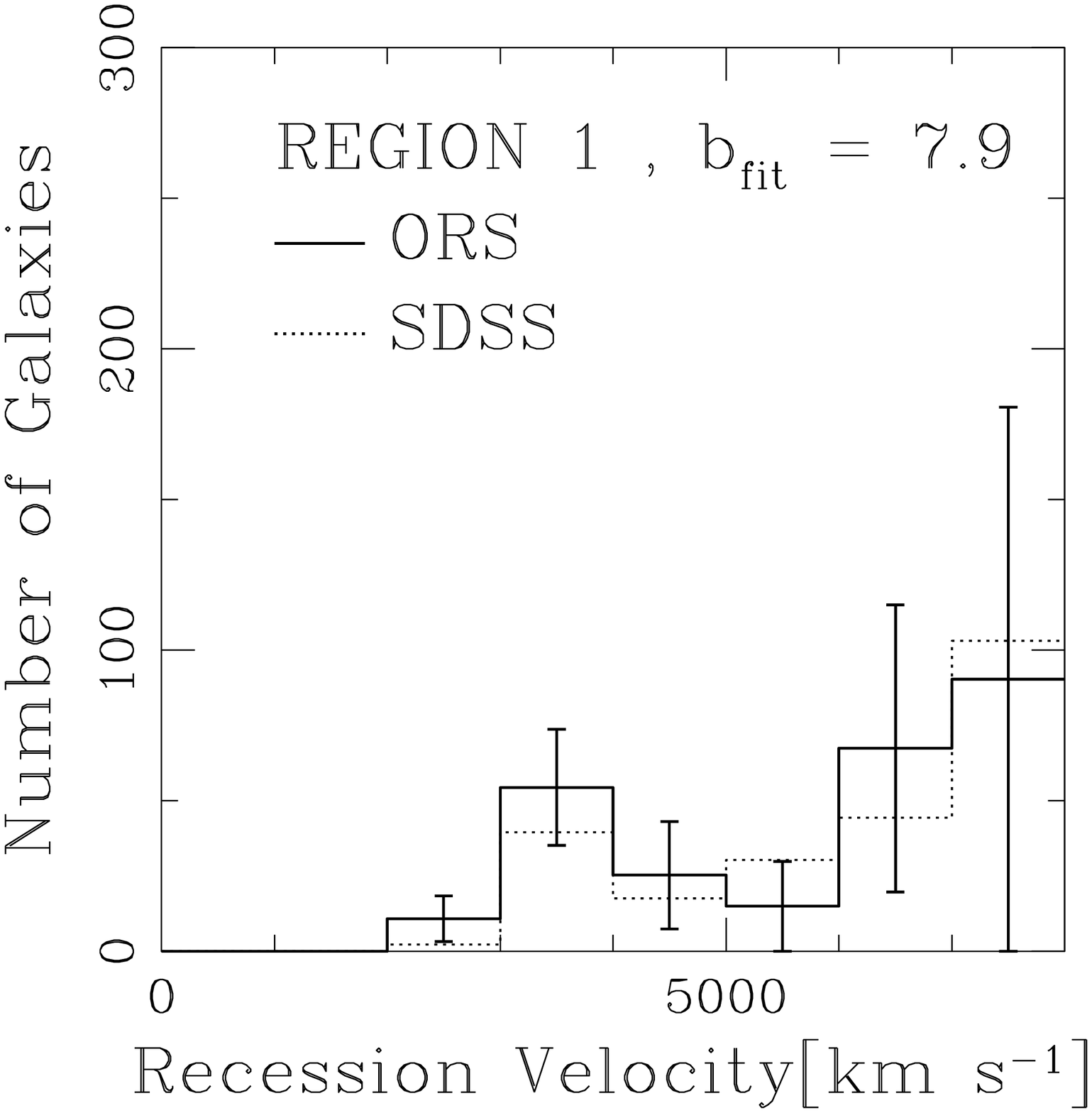}
	\FigureFile(40mm,40mm){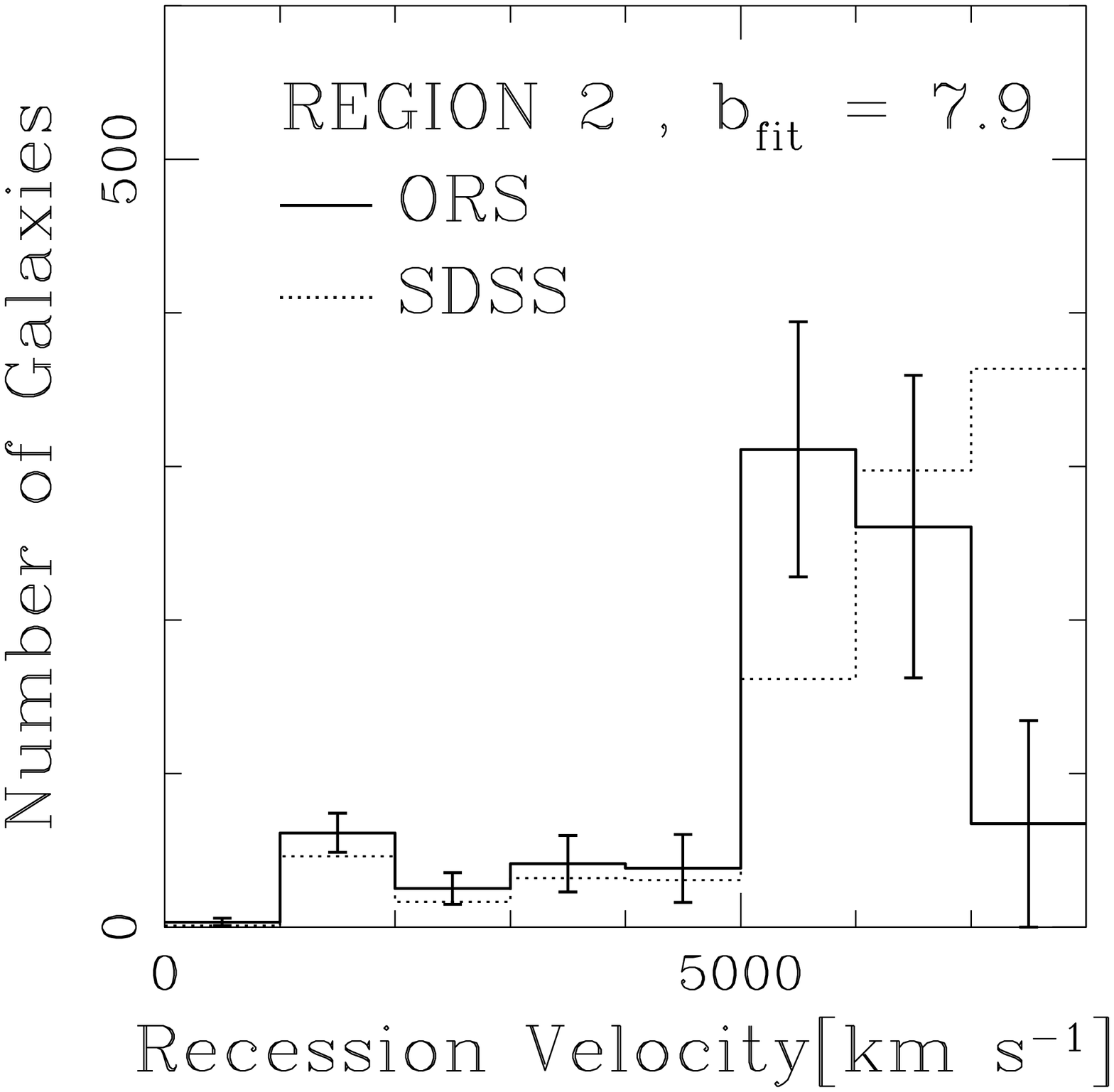}
	\FigureFile(40mm,40mm){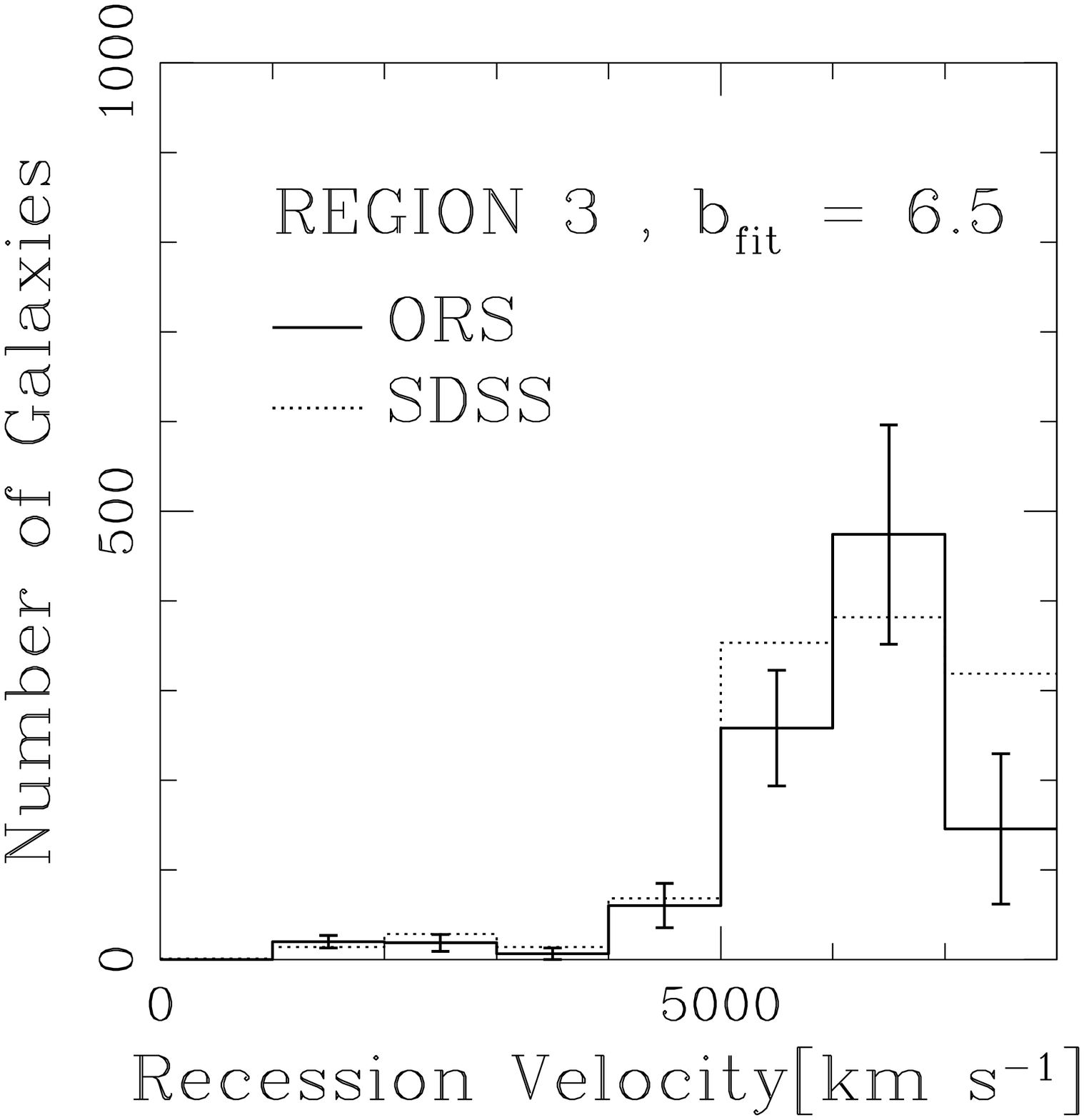}
	\FigureFile(40mm,40mm){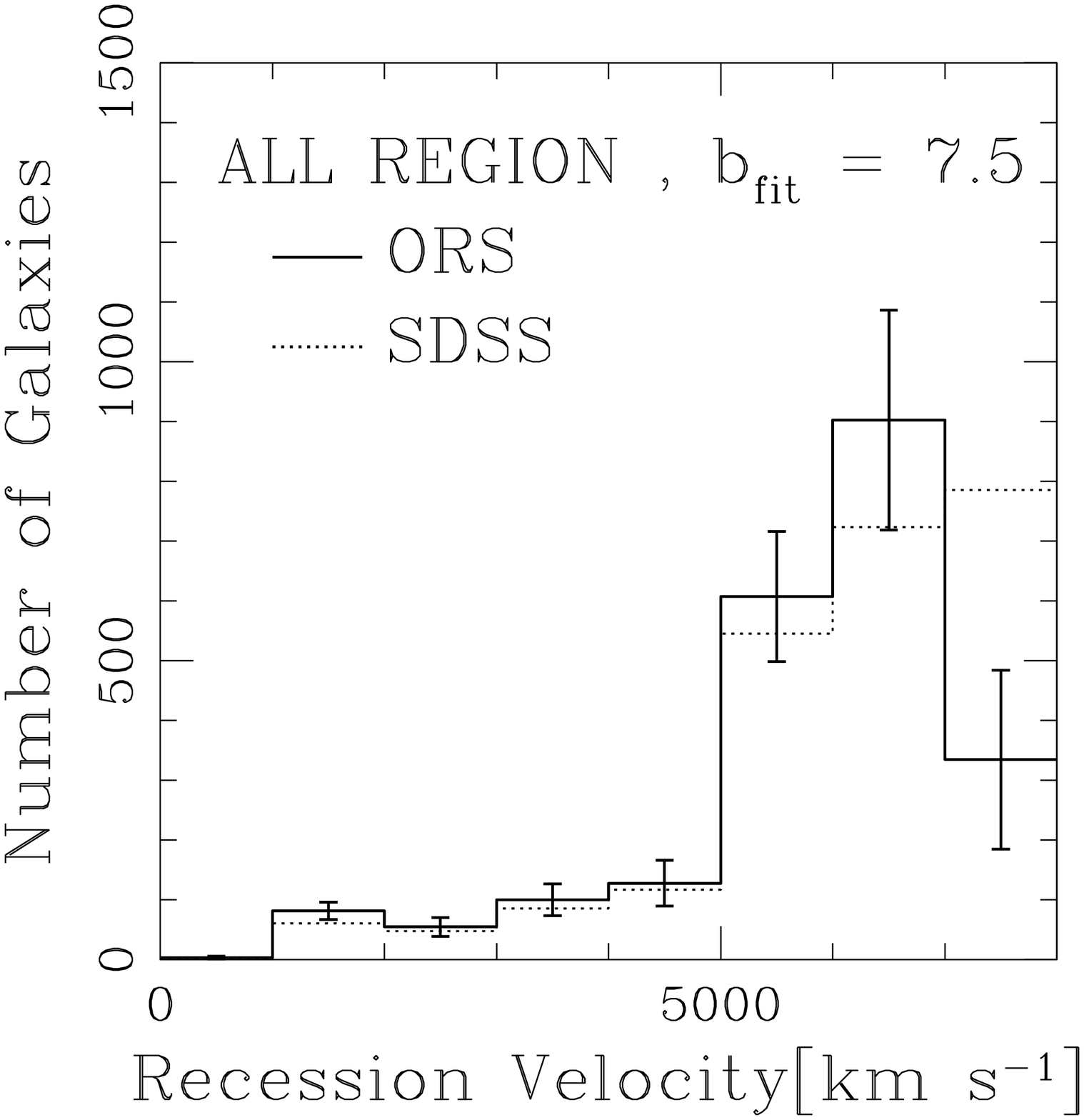}
\end{center}
\caption{
Number of galaxies in velocity bins of width 1000 km s$^{-1}$
in each region. Numbers from the SDSS are corrected
using the selection function derived in this study.
Numbers from the ORS are corrected by the selection function
given in Santiago et al.(1995) (upper panels)
and by the selection function revised in the present study (lower panels).
We show error bars on numbers of the ORS galaxies only, which are much larger
than those of the deeply sampled SDSS galaxies.
}
\label{fig8}
\end{figure}

\thispagestyle{empty}
\begin{figure}
\begin{center}
   \FigureFile(60mm,45mm){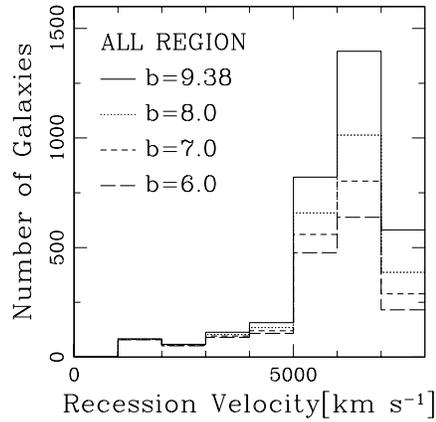}
\end{center}
\caption{
Number of ORS galaxies, which fall into the ALL region
and are corrected by the selection function given in Santiago et al.(1995),
for various choices of the parameter b.
It is clearly visible that the number counts at large distance
are sensitive to this parameter.
}
\label{fig9}
\end{figure}

\thispagestyle{empty}
\begin{figure}
\begin{center}
   \FigureFile(60mm,45mm){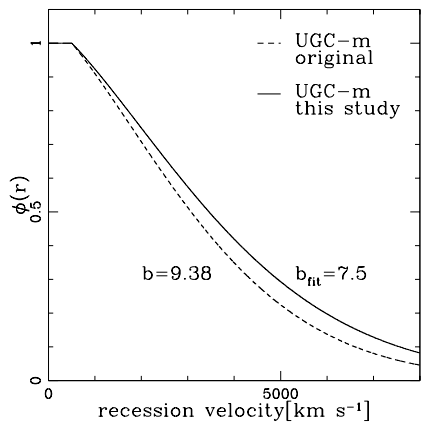}
\end{center}
\caption{
The selection function of the UGC-m with parameters
given in Santiago et al.(1995) (dotted line) and revised in this analysis
(solid line).
The revised $b$ value from the analysis in ALL region is used.
}
\label{fig10}
\end{figure}

\thispagestyle{empty}
\begin{figure}
\begin{center}
	\FigureFile(50mm,40mm){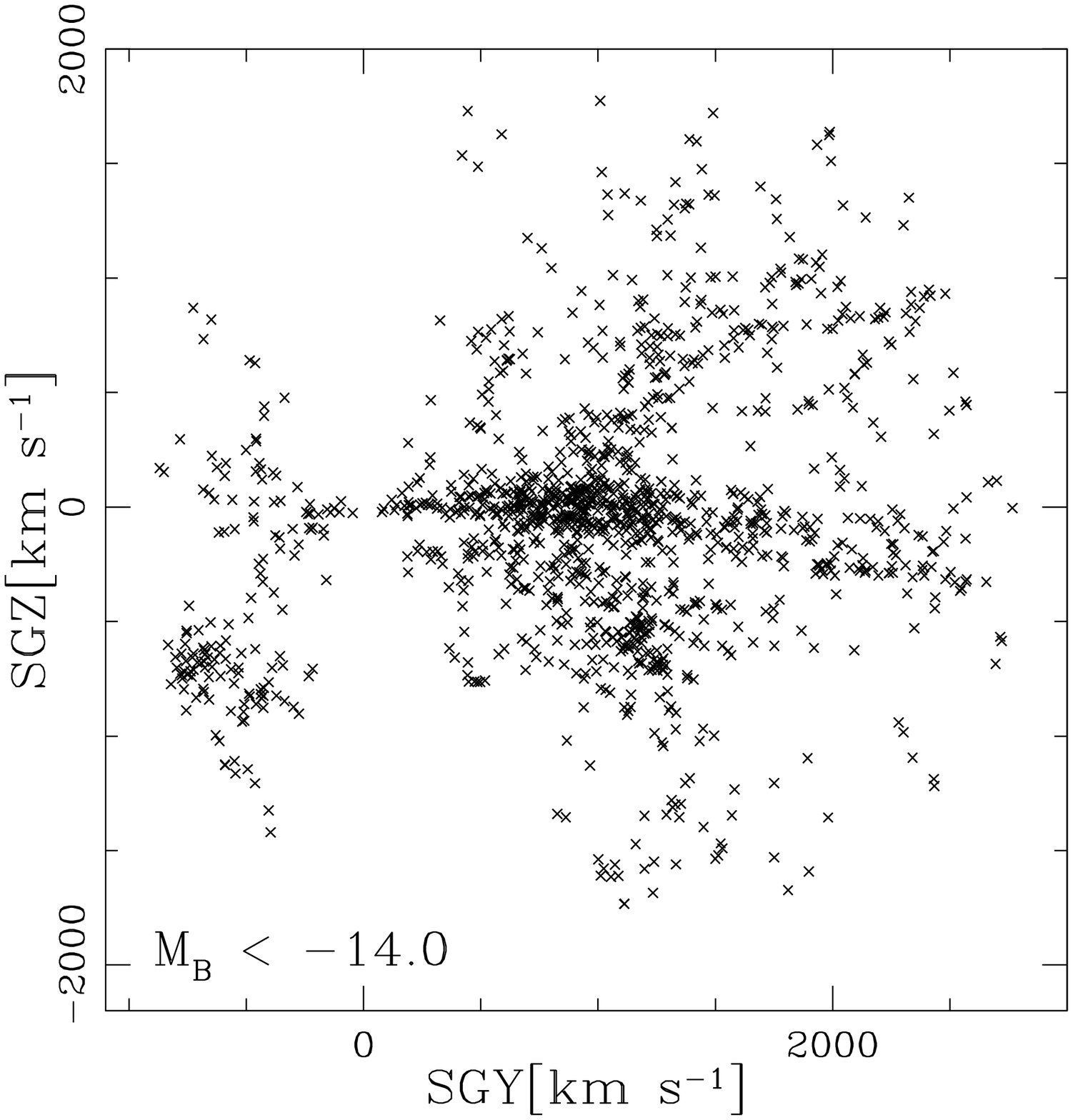}
	\FigureFile(50mm,40mm){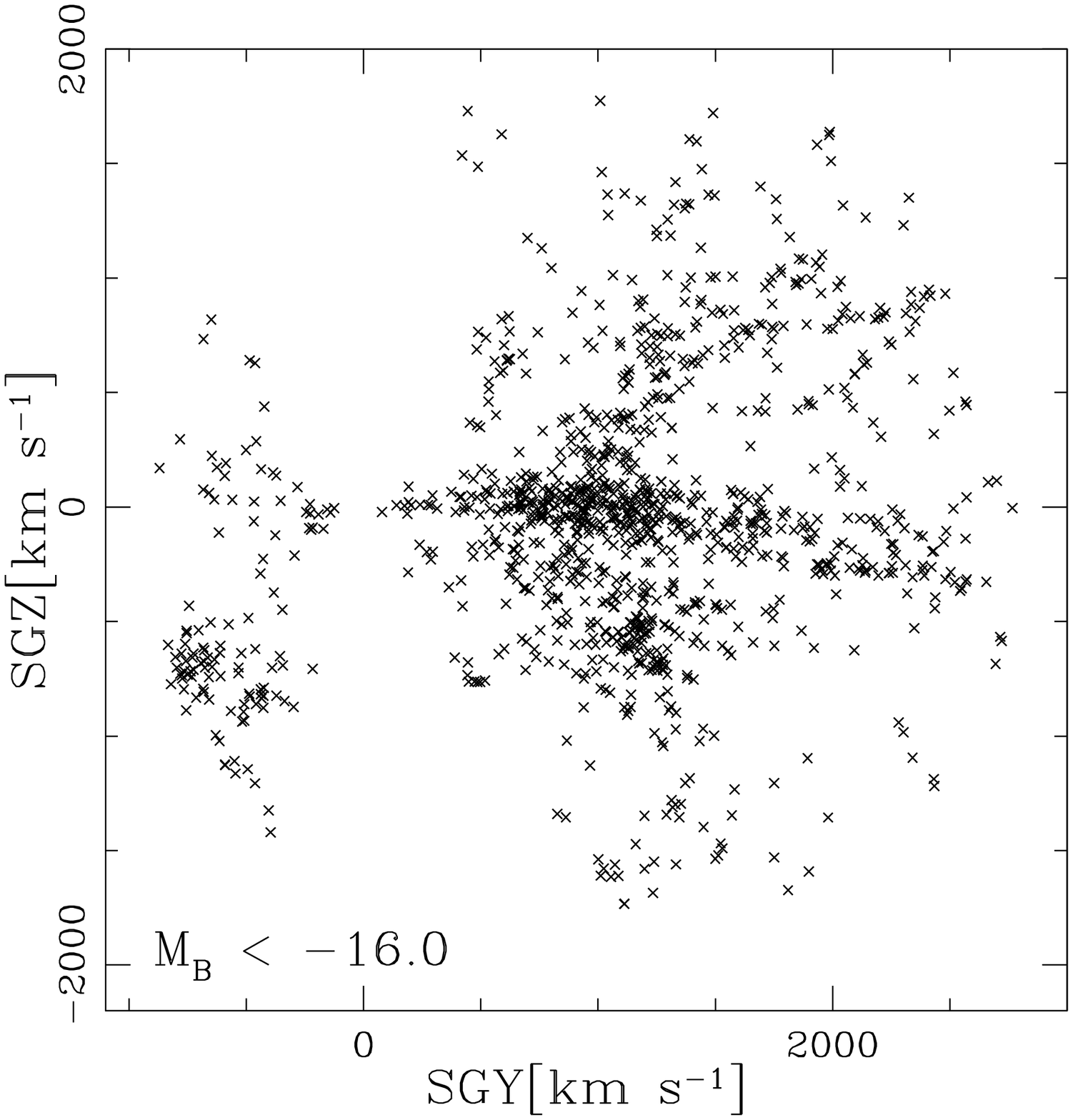}
	\FigureFile(50mm,40mm){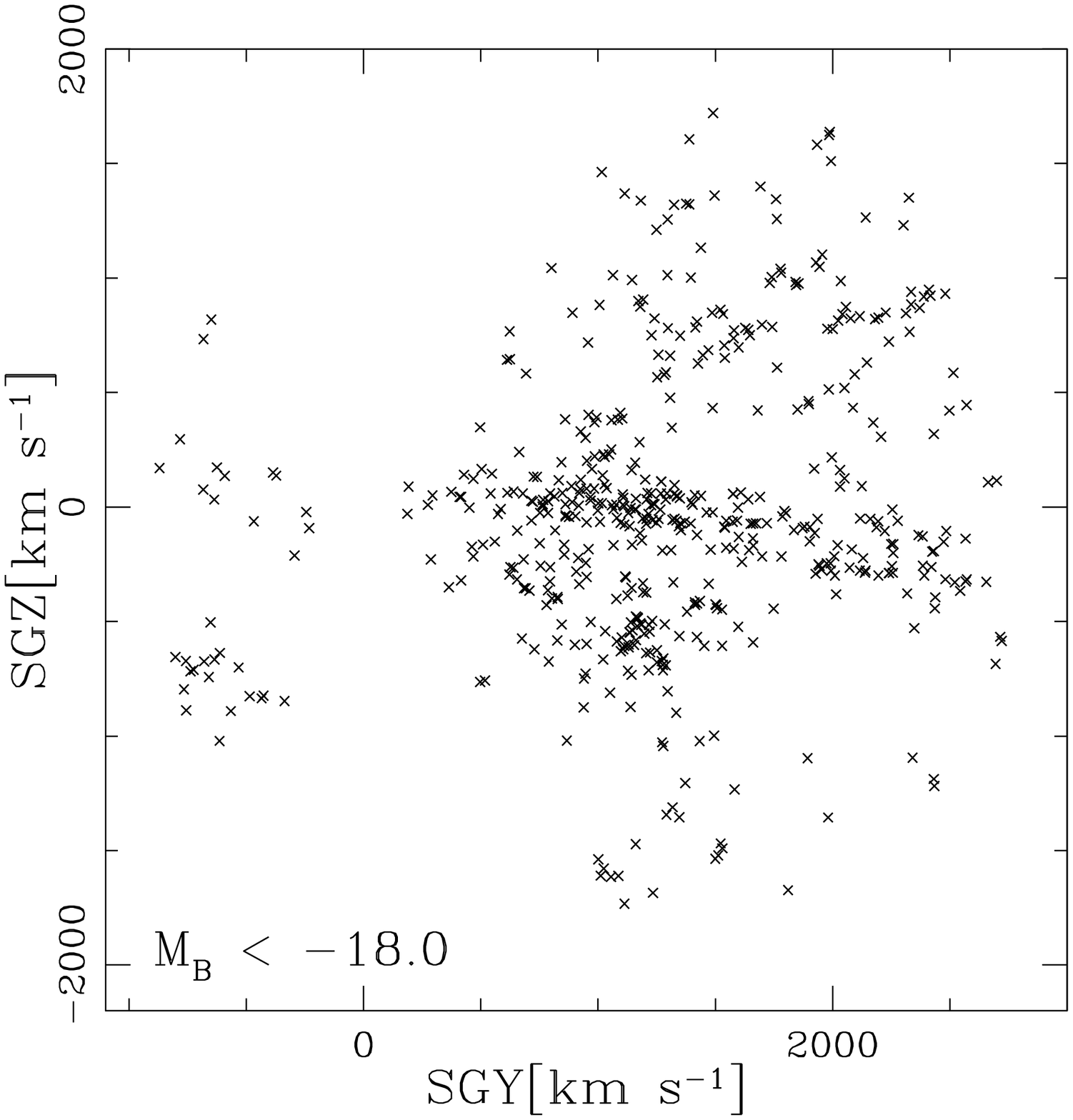}
\end{center}
\caption{
Galaxies more luminous than $M_{B}=-14,-16,-18$
in ESO-m and UGC-m data sample within 2000 km s$^{-1}$
from the Virgo cluster, which is the dense region
near SGZ=0 km s$^{-1}$, SGY=1000 km s$^{-1}$.
Our galaxy is located at the origin of the coordinate system.
}
\label{fig11}
\end{figure}

\thispagestyle{empty}
\begin{figure}
\begin{center}
	\FigureFile(70mm,60mm){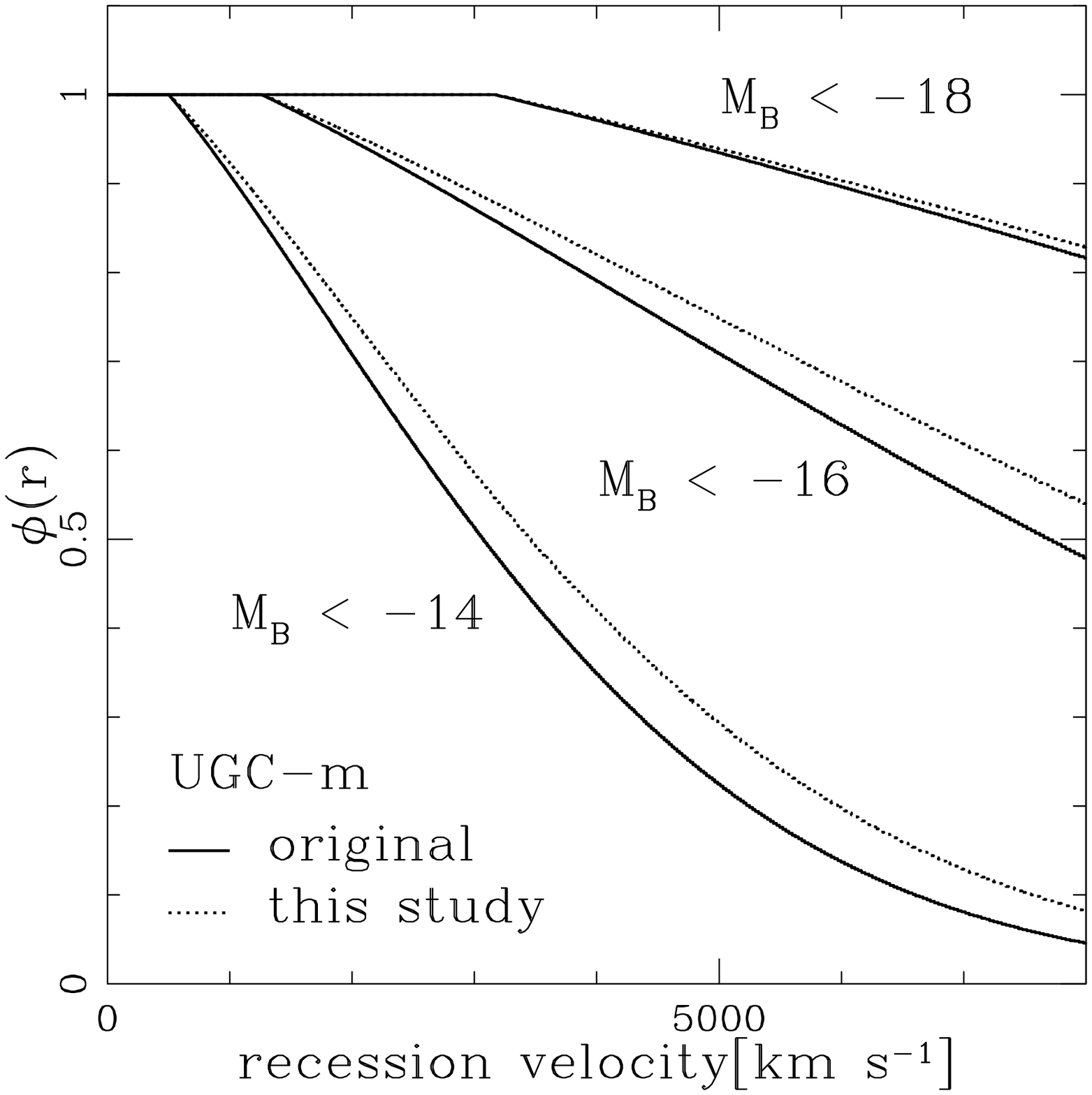}
	\FigureFile(70mm,60mm){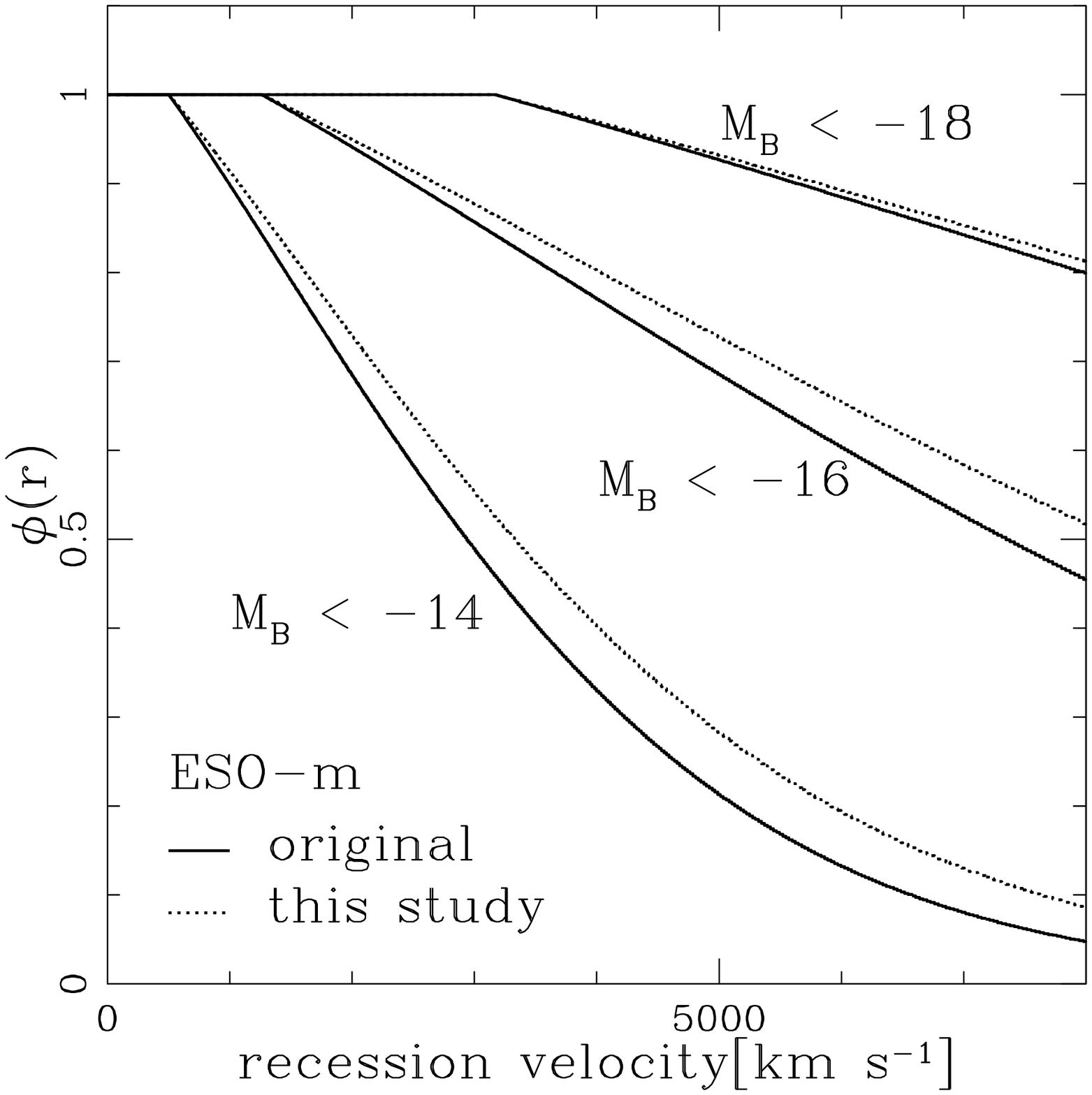}
\end{center}
\caption{
The selection function of UGC-m and ESO-m samples
for M$_{B} < -14,-16,-18$.
As for ESO-m data sample,
the revised $b$ parameter is calculated
so that the ratio of the value of the ESO-m selection function
with the revised $b$ value at 8000 km s$^{-1}$ to
that with the original $b$ value given in Santiago et al.(1995)
becomes equal to the ratio of the UGC-m selection function.
}
\label{fig12}
\end{figure}
\newpage

\thispagestyle{empty}
\begin{figure}
\begin{center}
	\FigureFile(50mm,40mm){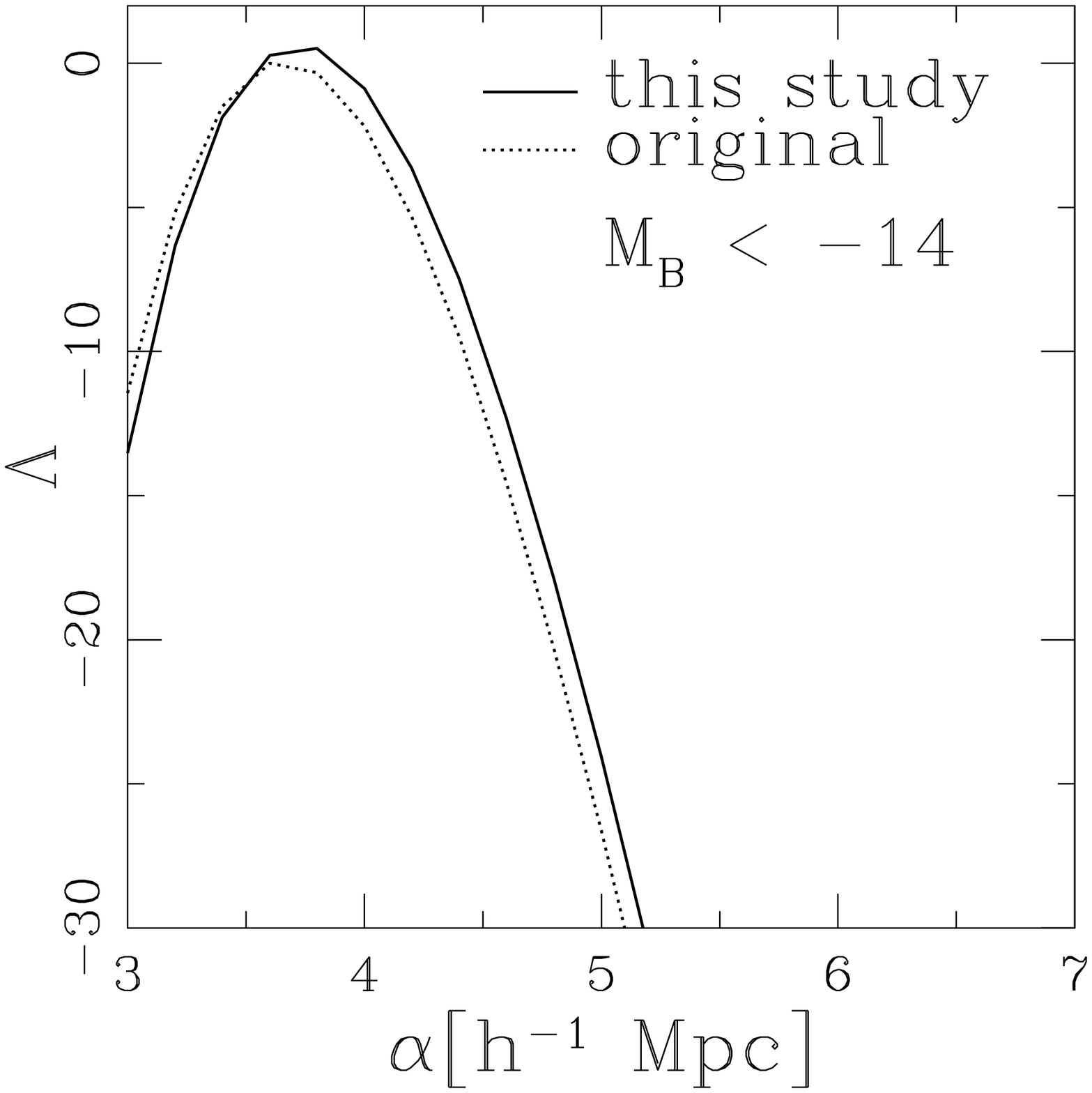}
	\FigureFile(50mm,40mm){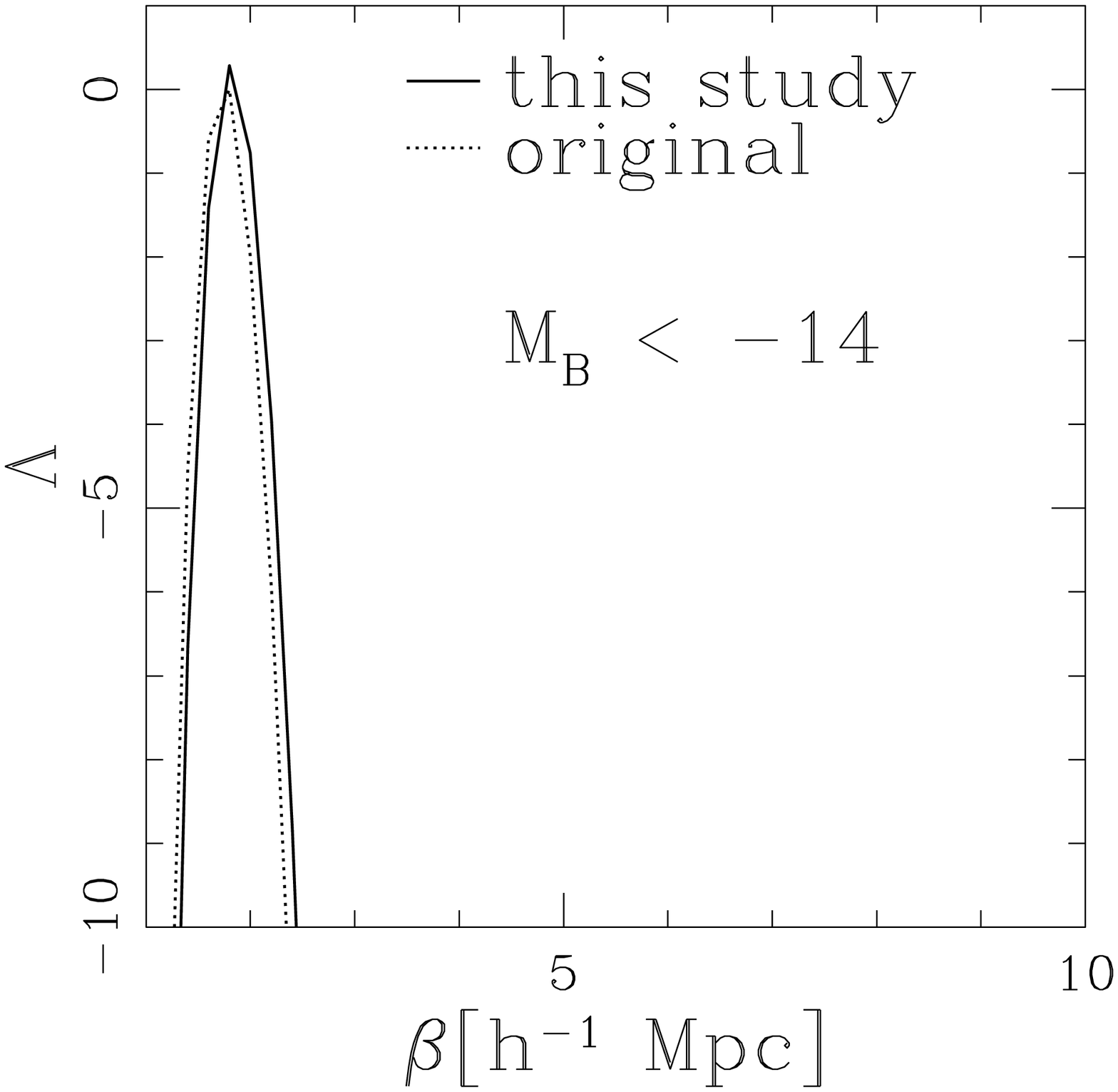}
	\FigureFile(50mm,40mm){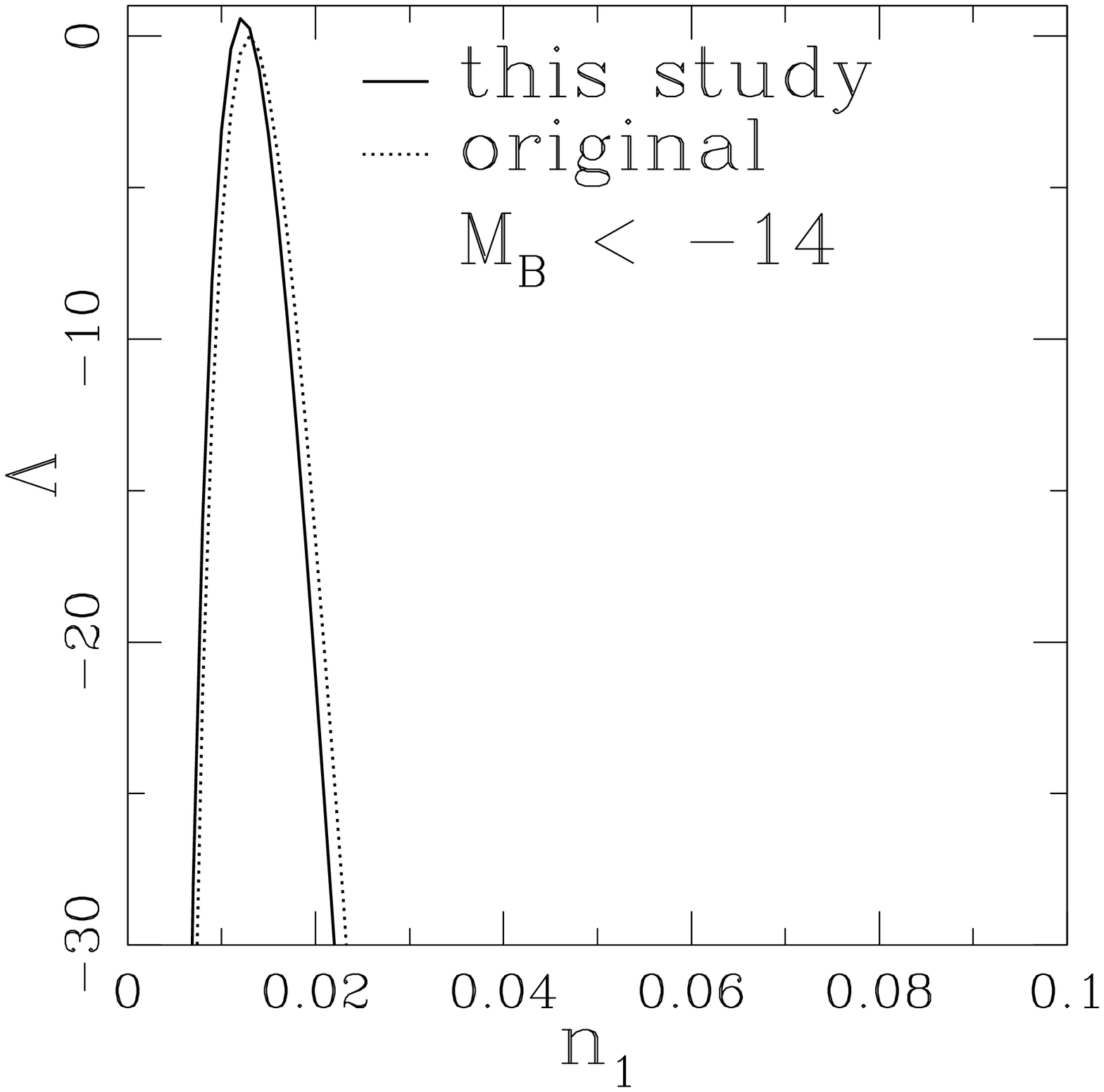}
	\FigureFile(50mm,40mm){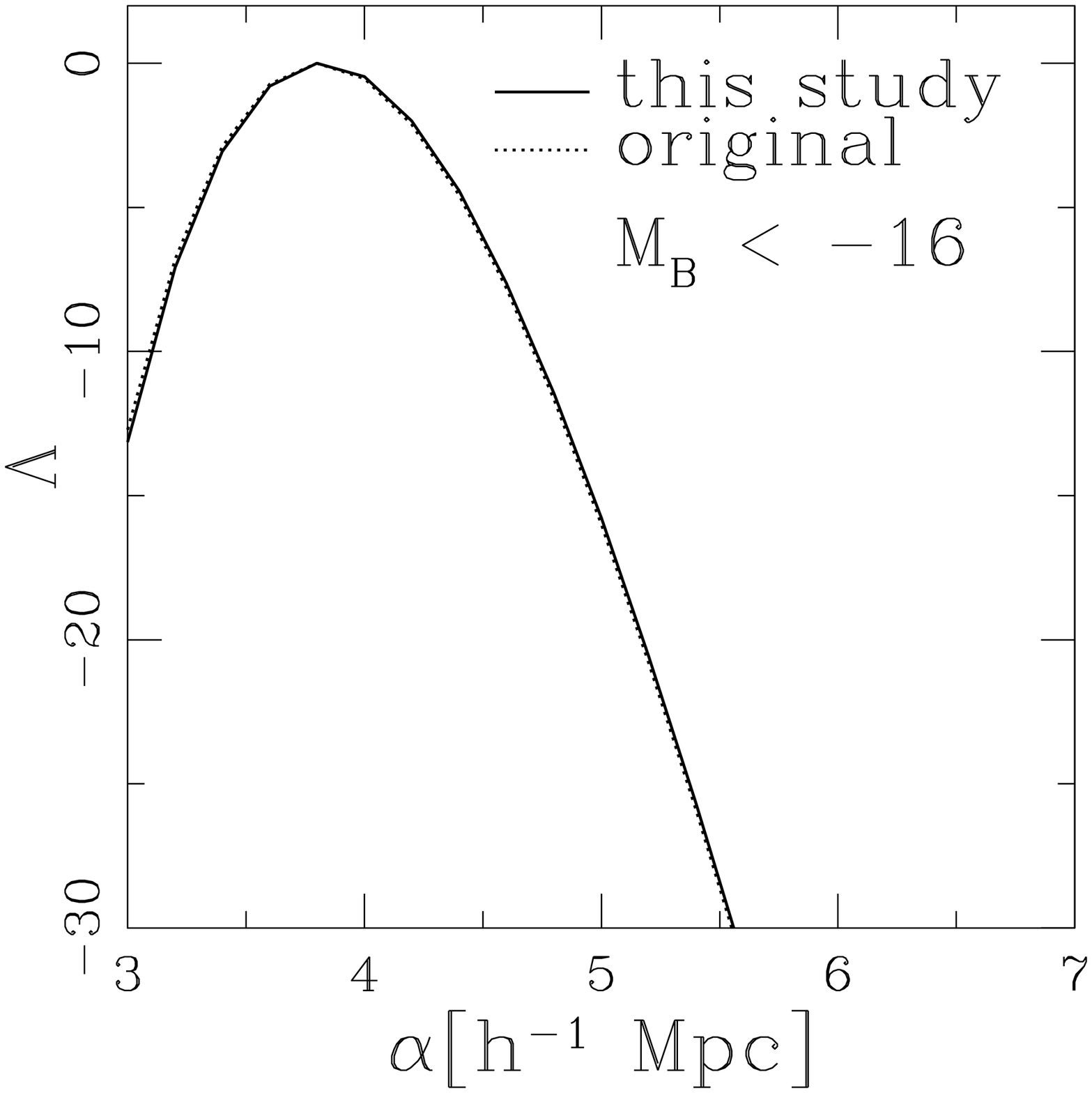}
	\FigureFile(50mm,40mm){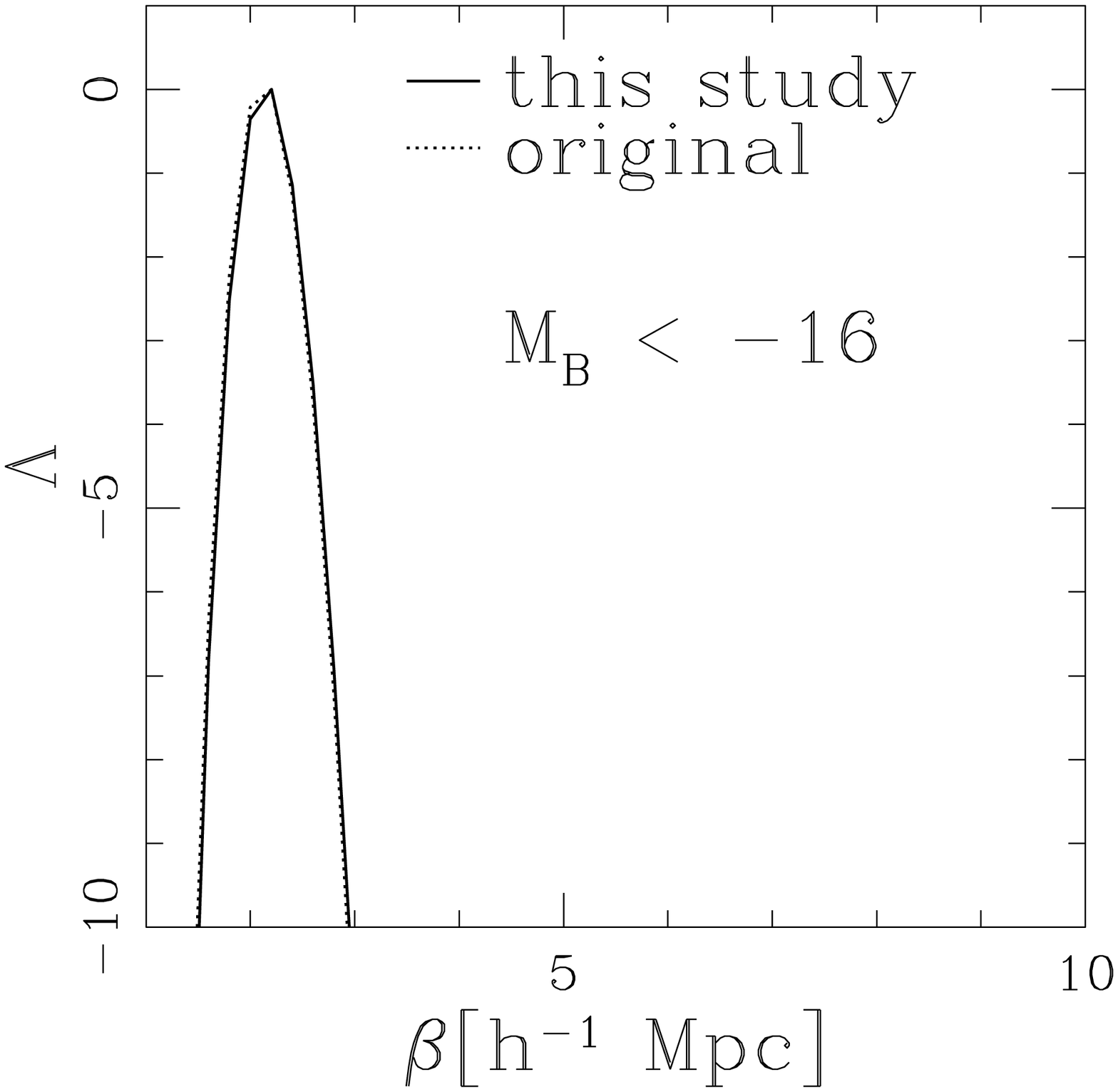}
	\FigureFile(50mm,40mm){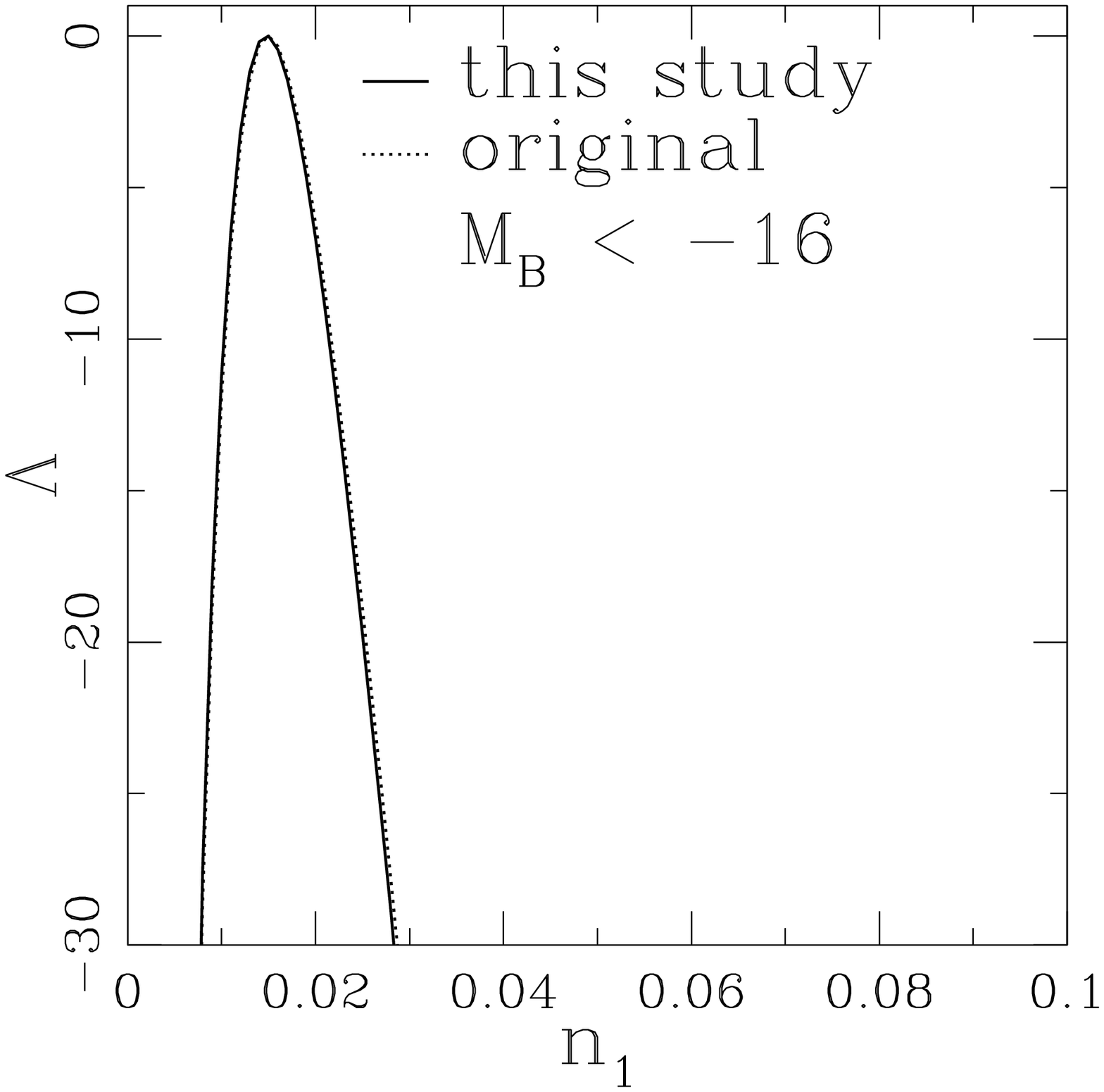}
	\FigureFile(50mm,40mm){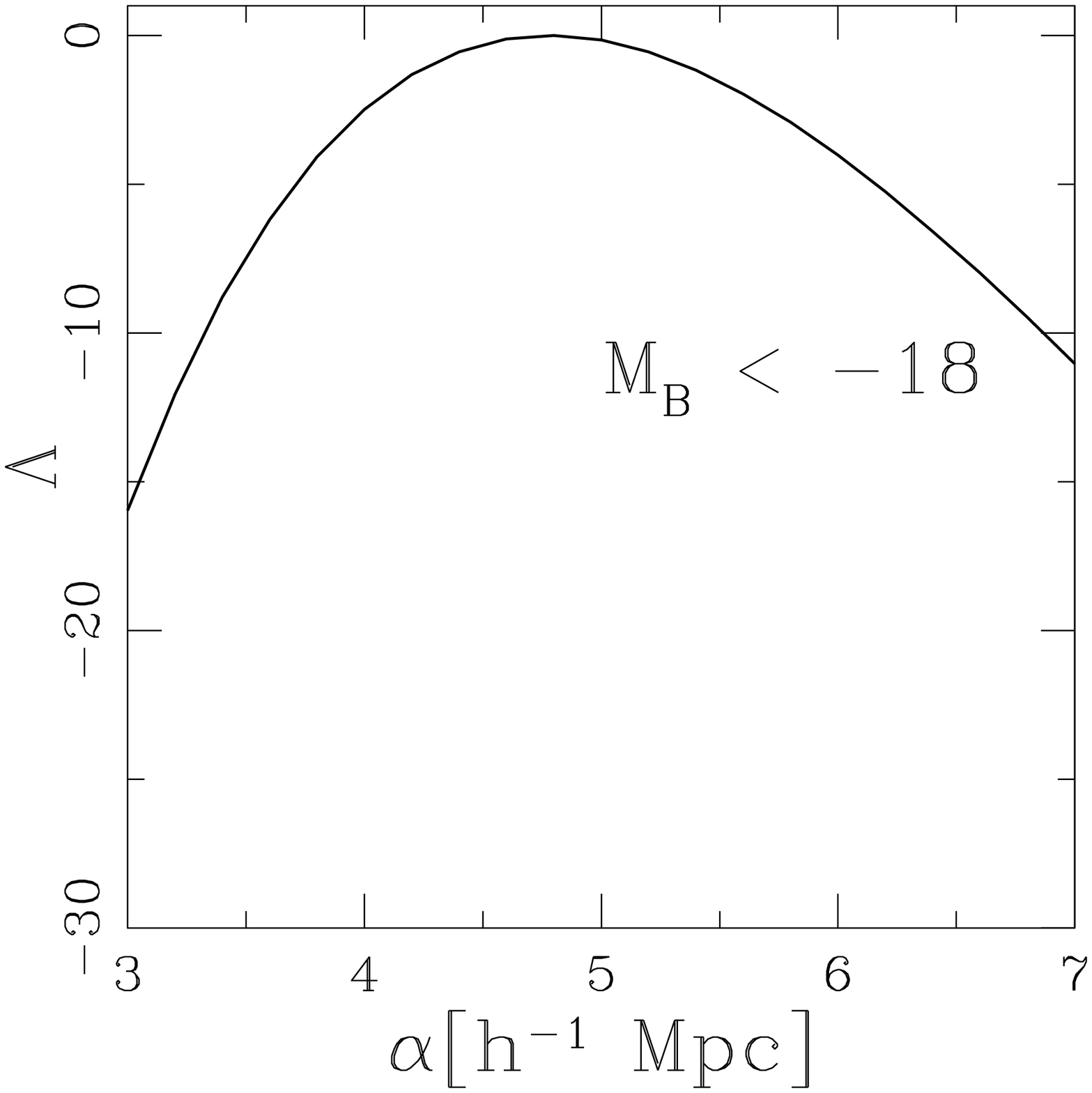}
	\FigureFile(50mm,40mm){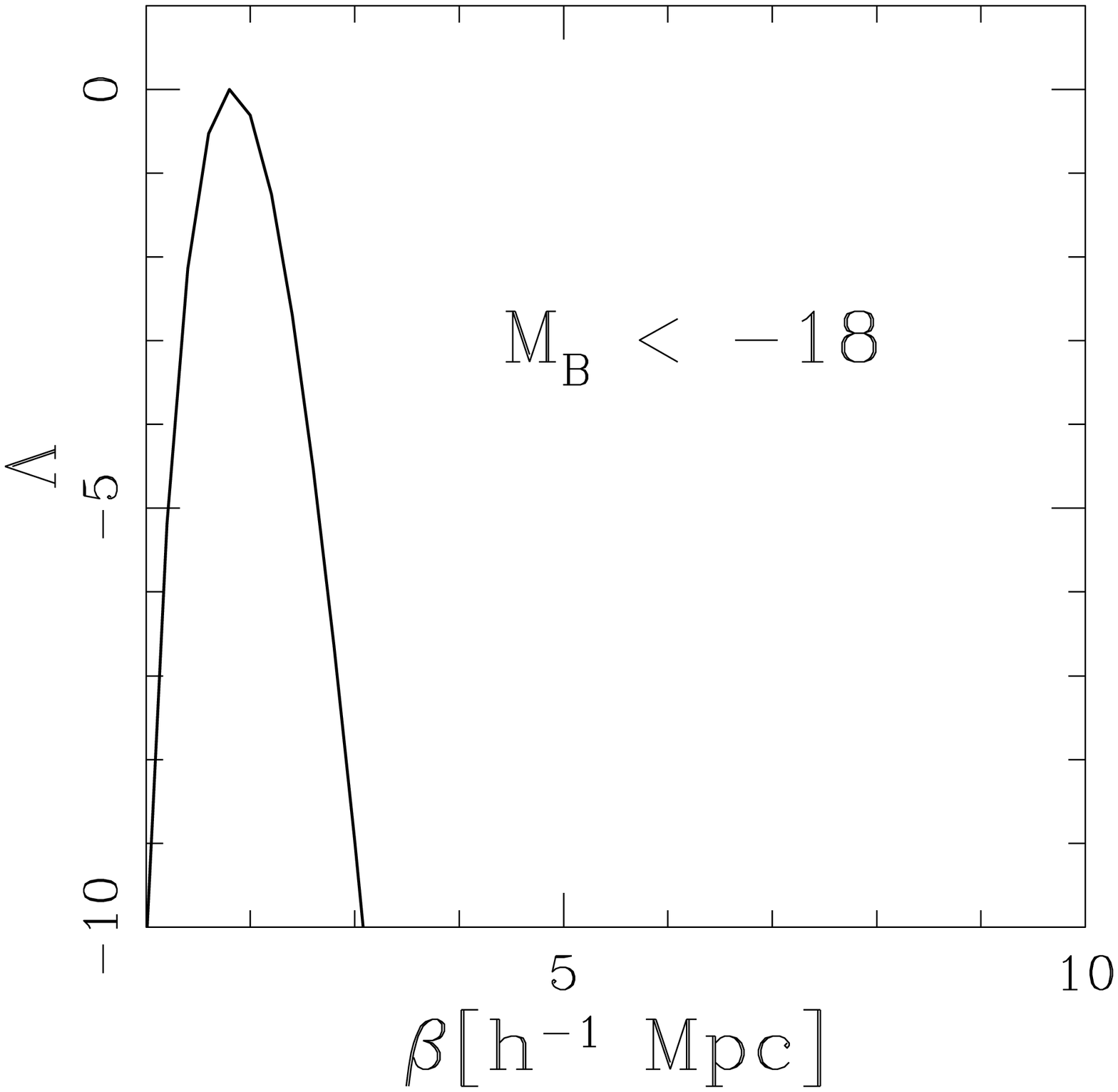}
	\FigureFile(50mm,40mm){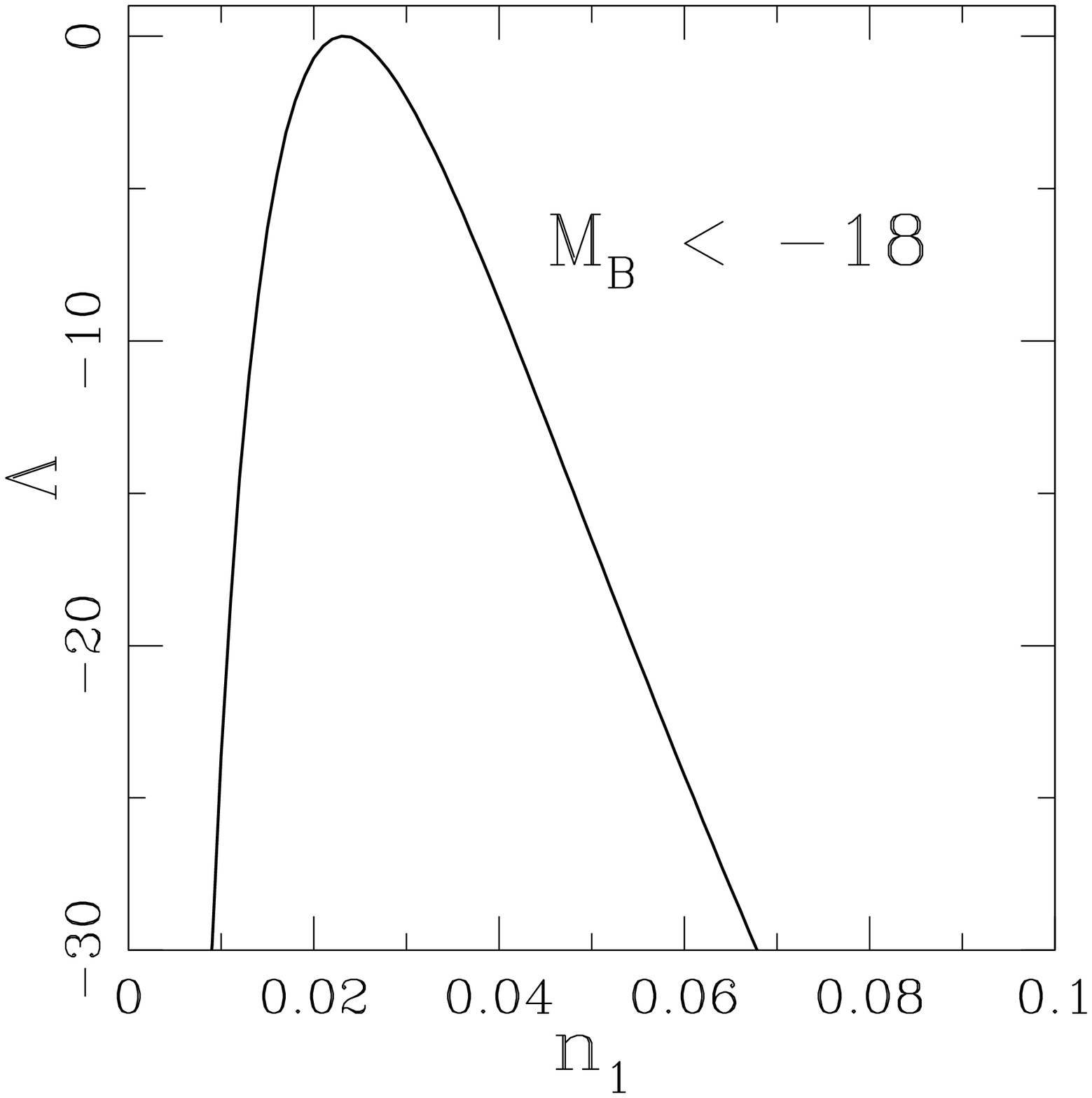}
\end{center}
\caption{
Values of the likelihood function with respect to $\alpha$ (left panels),
$\beta$ (center) and $n_{1}$ (right)
for luminosity range $M_{B}<-14$ (upper panels),
$M_{B}<-16$ (middle), $M_{B}<-18$ (lower).
The solid curves represents the results based on the revised
selection function while the dotted curves are based on the original selection
function.
Absolute values of the likelihood function
are normalized arbitrarily.
Note that for $M_{B}=-18$ both ESO-m and UGC-m
are complete samples in the limited-volume of radius
$\sim$ 2000 km s$^{-1}$ from the Virgo cluster (See Figure~\ref{fig2}).
}
\label{fig13}
\end{figure}
\newpage

\end{document}